\documentclass[12pt,letterpaper,super,comma,sort&compress, numbers]{article}
\usepackage[a4paper, total={7in, 10in}]{geometry}

\usepackage{graphicx}
\usepackage{helvet}
\usepackage{authblk}
\usepackage{hyperref}
\usepackage{amsmath} 
\usepackage{amssymb} 
\usepackage{orcidlink} 
\usepackage[super,comma,sort&compress]  
   {natbib}
\usepackage[right]{lineno} \linenumbers

\usepackage{amssymb}
\usepackage{amsmath}

\usepackage{siunitx}

\makeatletter
\renewcommand{\maketitle}{\bgroup\setlength{\parindent}{0pt}
\begin{flushleft}
  \textbf{\@title}

  \@author
\end{flushleft}\egroup}
\makeatother

\title{On-demand thermal power amplification enabled by active heat $Q$-switching}
\date{}

\author[1,$\dagger$]{Qian Ye}
\author[1,$\dagger$]{Aleida Machorro-Ortiz}
\author[1]{William Schmid}
\author[2]{Geoff Wehmeyer}
\author[1,3,4,**]{Naomi J. Halas}
\author[1,5,*]{Alessandro Alabastri}

\affil[1]{Department of Electrical and Computer Engineering, Rice University, Houston, TX, USA}
\affil[2]{Department of Mechanical Engineering, Rice University, Houston, TX, USA}
\affil[3]{Department of Physics and Astronomy, Rice University, Houston, TX, USA}
\affil[4]{Department of Chemistry, Rice University, Houston, TX, USA}
\affil[$\dagger$]{These authors contributed equally}
\affil[5]{Lead contact}
\affil[*]{Correspondence: alabastri@rice.edu}
\affil[**]{Correspondence: halas@rice.edu}

\begin{document}
\nolinenumbers
\maketitle

\section*{SUMMARY}

Thermal management underpins essentially every energy technology, from solar harvesters and waste-heat recovery to industrial process heating and electronic cooling. Yet, thermal systems remain limited compared to their electrical and optical counterparts: they lack a direct equivalent of active control elements that enable on-demand pulse generation. As a result, in conventional thermal energy storage architectures, the amount of energy stored and the peak power at which it can be released are both fixed at design time by material properties and heat-exchanger geometry, locking each device to a single operating point in the energy--power plane. This rigidity is incompatible with applications that require short, high-power thermal bursts on demand. Here we show that a counter-flow heat oscillator admits an actively switchable effective thermal quality factor, $Q$, and that modulating $Q$ on sub-dwell-time scales generates transient outlet power exceeding the steady input by more than an order of magnitude. We formalize the system as a dissipative resonant thermal cavity with $Q$ controlled by the balance between advective and conductive transport and environmental losses, and we experimentally demonstrate, in a water-based dual-channel device, $\sim$5-fold transient power amplification through controlled flow detuning, in quantitative agreement with our thermofluidic model. Using the validated model, we predict that liquid-metal implementations achieve $\sim$20$\times$ peak amplification, while vertically coupled oscillators reach $\sim$40$\times$. Unlike static sensible, latent, or thermochemical storage media that occupy fixed operating points, an actively switched heat oscillator traces continuous trajectories through the Ragone plane, separating energy harvesting from energy delivery as an independent control axis. Active $Q$-switching thereby establishes a distinct mode of thermal power management, accessing $\sim$5$\times$ (demonstrated experimentally) to $\sim$40$\times$ (projected numerically) peak-power amplification on continuous input through a single architecture, a regime inaccessible to passive thermal storage and a missing analogue of the active pulse-generation tools long available in optics and electronics.

\section*{KEYWORDS}


active thermal control, $Q$-switching, resonant heat transfer, non-equilibrium heat transport, thermal metamaterials, photothermal systems

\section*{INTRODUCTION}

Thermal energy storage (TES) is a practical approach to offsetting intermittency in renewable heat sources and valorizing industrial waste heat.\cite{Avghad2016} Among TES approaches, sensible heat storage---storing energy in water, rocks, oils, or other abundant media---remains attractive because of its material availability, safety, straightforward manufacturability, and low cost per unit energy stored.\cite{Sharma2025} These features are especially important for decentralized deployments, where supply-chain constraints, maintenance, and system simplicity can dominate over absolute round-trip efficiency.\cite{Raghav2020}

A fundamental limitation of conventional TES architectures, whether sensible, latent, or thermochemical, is that their operating point in the energy--power plane is fixed by material choice and heat-exchanger geometry. Heat is stored in static media or slowly circulating loops, and discharge is constrained by the thermal conductance of heat exchangers, allowable temperature approaches, and the piping and pumping infrastructure, which is typically sized for peak delivery. A TES system may therefore store substantial energy yet be unable to deliver short, high-power thermal bursts without oversizing hardware or adding electro-thermal conversion stages \cite{Saleem2024}. Many relevant loads, however, are inherently transient or batch-like (e.g., sterilization/cleaning cycles,\cite{Subekti2011} rapid distillation start-up,\cite{Reepmeyer2004} thermal swing adsorption/regeneration,\cite{GomezRueda2022} microfluidic systems\cite{Miralles2013} or controlled reactor ramps\cite{Luyben2019}), motivating platforms that can both accumulate energy efficiently and release it rapidly and programmably.\cite{Hurst2024,Woods2021}

Here, we introduce thermal Q-switching as a fundamentally different paradigm for controlling heat flow. By exploiting resonant energy transfer (RET) in coupled fluidic channels, thermal energy is accumulated through internal recirculation characterized by a high effective quality factor $Q$, and subsequently released on demand by switching the system to a low-$Q$ state. This transition enables the rapid conversion of stored enthalpy into a heat pulse, with output transient power exceeding the steady input ($P_{out}>P_{in}$) consistent with energy conservation. Unlike conventional TES or thermal switching strategies, this approach provides time-domain programmability of thermal power, analogous to Q-switching in photonic systems or pulse generation in electrical circuits. As a result, thermal Q-switching establishes a new degree of freedom in thermal energy systems, enabling dynamic access to operating regimes that are inaccessible to static storage or diffusive transport architectures.

The notion of controlling energy retention and release by switching an effective quality factor is widely used in other resonant systems, most notably $Q$-switched lasers, where intracavity losses are transiently increased and then rapidly reduced to release stored energy as an intense optical pulse.\cite{Svelto2013} Here we adopt a similar control principle: high-$Q$ charging followed by low-$Q$ discharge, implemented through fluidic detuning or interfacial thermal decoupling rather than optical loss modulation. While detuning can be achieved by flow-rate modulation, thermal decoupling can be achieved using active heat switches that modulate the thermal resistance at the fluid interface.\cite{Wehmeyer2017,Ram2023,Kommandur2023} Because discharge is driven by releasing previously stored energy rather than by increasing input power, the transient outlet power can exceed the steady input power.

Research on wave-like and non-equilibrium heat transport has advanced substantially in recent years,\cite{Li2019,Li2021,Qi2022,Zhang2023,Ju2023,Yang2024,Liu2024,Cao2024,Wang2026} and thermal systems with interacting moving fluids have been shown to exhibit oscillator-like dynamics.\cite{Alabastri2020}. Prior work established that counter-flow heat oscillators can recirculate thermal energy at a tunable \emph{steady-state} operating point, characterized by a figure of merit that quantifies internal heat recycling.\cite{Alabastri2020,Ye2023} 
Those studies, however, addressed the stationary regime: the oscillator was characterized at its steady-state resonant operating point, where $Q$ is a fixed property of the geometry and flow conditions and the outlet power is bounded by the steady input. The present work instead addresses the time-dependent response of the same class of systems, where switching $Q$ on sub-dwell-time scales gives access to transient performance --- on-demand release and outlet power far exceeding the steady input --- that has no counterpart in stationary systems.
The central result of the present work is that this recirculation can be \emph{actively switched} on a sub-dwell-time scale, producing transient outlet power that exceeds the steady input by more than an order of magnitude. Specifically, in this work, we (i) formalize the oscillator as a dissipative resonant cavity with an explicit, predictive $Q$-factor; (ii) demonstrate experimentally and numerically that $Q$ can be switched between high-retention and fast-release states through two independent, physically distinct actuation modes (flow detuning and interfacial decoupling), without modifying the input power; and (iii) show that the resulting system traces continuous trajectories through the thermal energy density - thermal power density Ragone space, rather than occupying a fixed point as static TES media do. This conceptual shift, from passive storage to actively controllable non-equilibrium thermal transport, is what enables outlet transients $\gg P_\mathrm{in}$ without violating energy conservation.

\begin{figure}[htbp]
  \centering
    \includegraphics[width=0.7\textwidth]{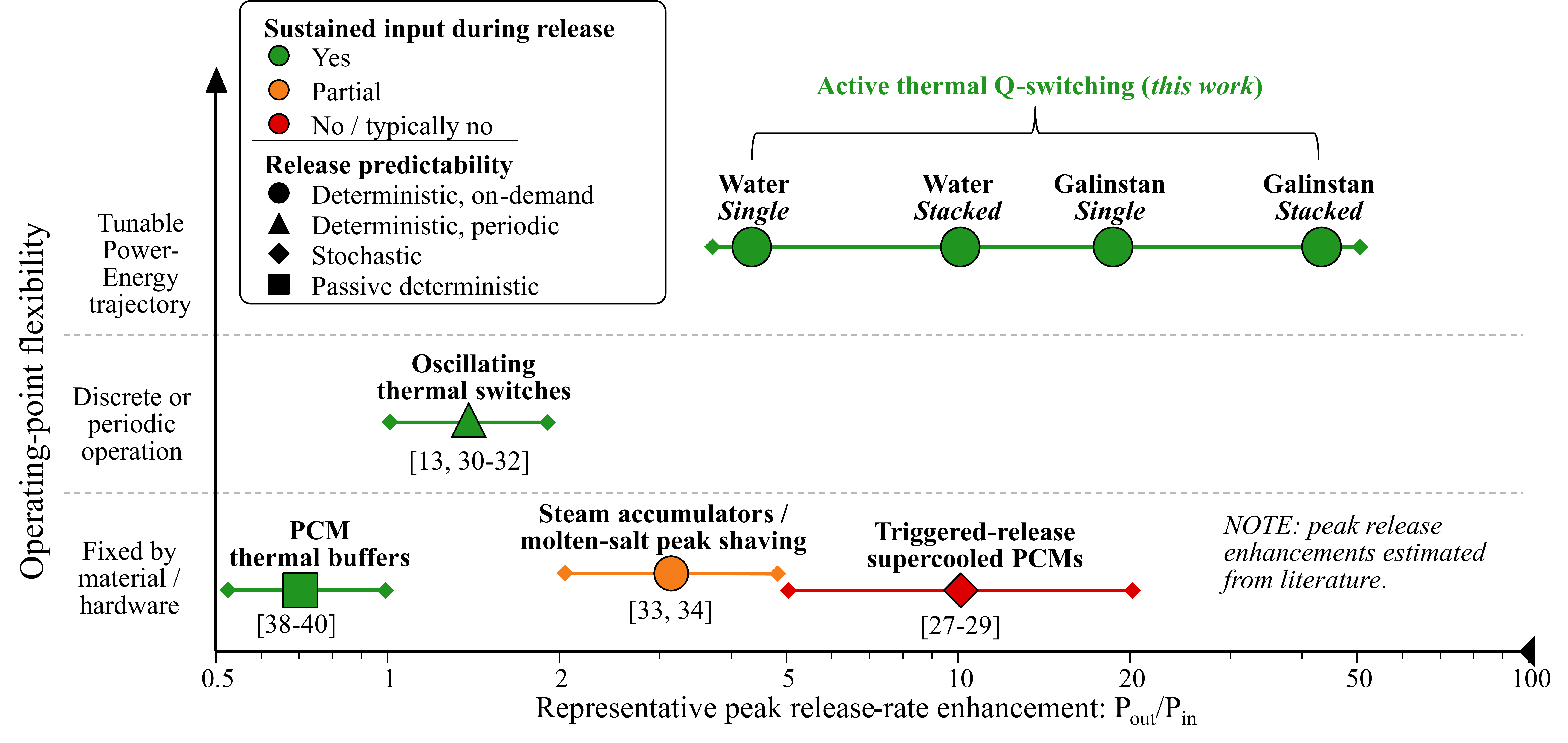}
  \caption{
\textbf{Approaches to transient heat generation from static thermal input.} Representative approaches are mapped by estimated peak release-rate enhancement, $P_{out}/P_{in}$, and operating-point flexibility. Marker color indicates whether the input can remain on during release, and marker shape indicates release predictability. Active thermal Q-switching occupies a distinct region by combining deterministic on-demand release, sustained-input compatibility, and tunable power–energy trajectories within the same hardware. Values for prior methods are representative estimates from the cited literature.}
  \label{fig:comparison}
\end{figure}

\textbf{Advance over the state of the art.} The problem of producing transient, high-power thermal output from a steady-state heat source is not new. Three families dominate the prior art. First, \emph{triggered-release supercooled phase-change materials} (PCMs), most prominently sodium acetate trihydrate and related salt hydrates and polyols, are charged by heating to melt the storage medium, then held in a metastable supercooled liquid state until nucleation is triggered mechanically, electrically, ultrasonically, or by seeding.\cite{Beaupere2018,Chen2023,Wang2025AMPD} Second, \emph{oscillating thermal switches} \cite{Zhu2022OscillatingGadolinium}, placed between a constant heat source and a downstream heat engine alternate between high- and low-resistance states, maintaining a maximum temperature gradient across the engine and improving harvesting power and efficiency relative to constant-flux operation.\cite{McKay2013ThermalPulse,Wehmeyer2017,Chen2015TEG} Third, \emph{steam accumulators and molten-salt peak-shaving systems} release prior sensible or latent storage, typically by depressurization or controlled mixing.\cite{Forsberg2017LWR,Wei2026MoltenSalt} Each of these approaches, while addressing a need, has structural limits that distinguish it from the present work. Supercooled PCMs require a solid--liquid phase transition every cycle, and their release timing depends on nucleation kinetics that remain a reliability challenge in the field; the recharge time is set by the latent heat of the material and is not a design degree of freedom. Oscillating thermal switches do produce thermal pulses through cyclic store-and-release in the source's thermal mass, but the architecture is optimized around forced periodic operation at a switch-defined frequency to maximize the harvested average power of a downstream engine, not around aperiodic, on-demand single-burst delivery whose amplification and timing are independently programmable. Industrial peak-shaving systems can release prior storage on demand and modulate output power against grid load, but their peak-power capacity and energy capacity are independently set at the build stage by heat-exchanger and turbine sizing and the trade-off between the two is not a control parameter. By contrast, active $Q$-switching of a resonant heat oscillator (i)~admits a range of thermal quality factors that maps to the achievable peak-power amplification; (ii)~operates in a single fluid phase without phase transitions or material consumption, eliminating the cycling-degradation modes that affect PCM-based pulse generators; (iii)~triggers release deterministically through a physical actuation (flow detuning or interfacial decoupling) with predictable pulse shape rather than through stochastic nucleation; and (iv)~traces continuous trajectories through Ragone space using the same hardware via control of $Q$. A comparison across the performance and operational axes is provided in Fig.~\ref{fig:comparison} and further details can be found in Table S1 within the Supplemental Information.

\section*{RESULTS AND DISCUSSION}

\subsection{Resonant energy transfer (RET) and $Q$-switching working principle}

\begin{figure}[htbp]
  \centering
    \includegraphics[width=0.7\textwidth]{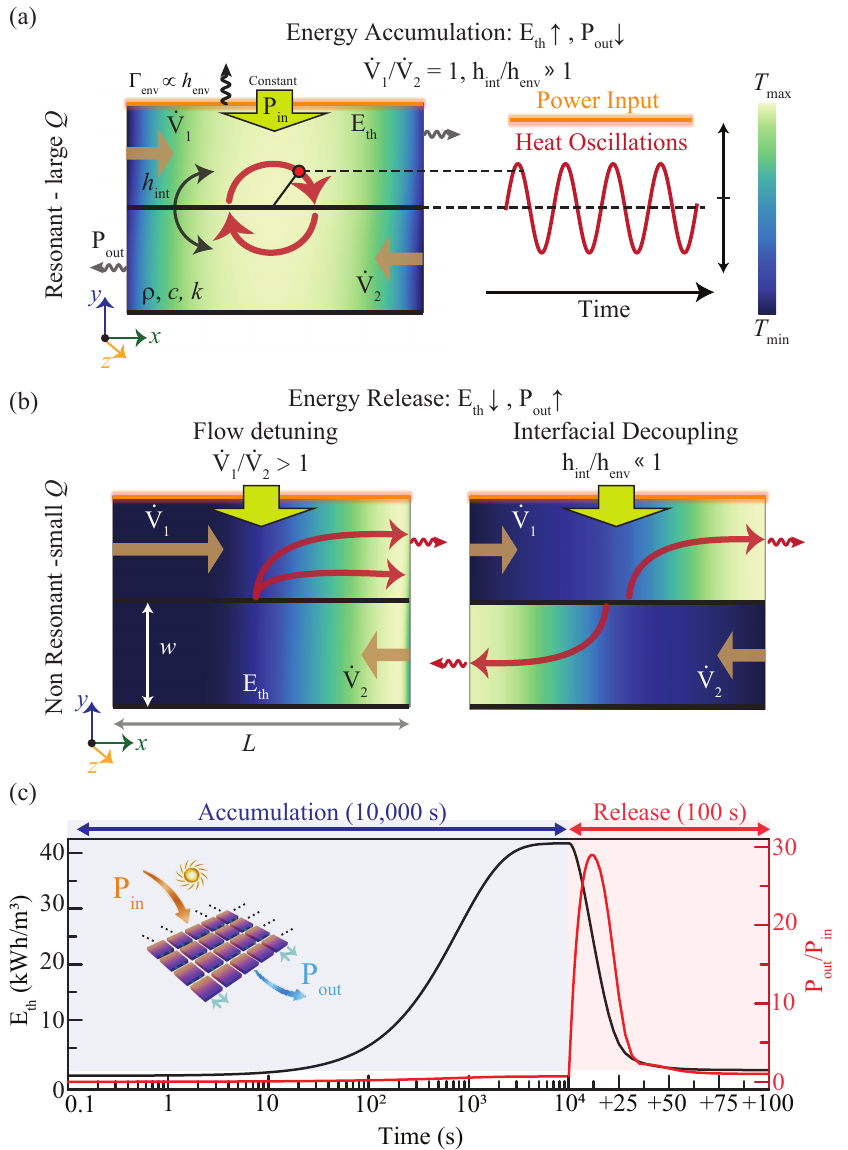}
  \caption{
\textbf{Q-switching for thermal power amplification (concept and simulations).}
\textbf{a)} Resonant Configuration. Two countercurrent flow rates, $\dot{V}_1$ and $\dot{V}_2$, with density $\rho$, specific heat $c$, and thermal conductivity $k$, exchange heat through a thermal conductor at the interface (effective heat transfer coefficient $h_{\mathrm{int}}$) of channels with length $L$ and thickness $w$. A steady input power $P_{in}$ enters the top surface of fluid 1. The two-channel system loses power partially to the environment, $\Gamma_{\mathrm{env}}$, and through the outlets, $P_{\mathrm{out}}$. When $\dot{V}_1/\dot{V}_2 = 1$, the system is at resonance, it sustains large heat oscillations, accumulates thermal energy $E_{\mathrm{th}}$, and exhibits a large $Q$-factor. 
``Heat oscillations'' here are understood in the dynamical-systems sense: they describe the spatiotemporal trajectory of the heat flux exchanged across the channels. Circling red arrows on the left represent heat flow in space. The harmonic red line on the right shows the dynamics of heat flow between the channels over time. 
\textbf{b)} Non-Resonant Configuration (short time after $Q$ switching). \emph{Left:} same as in (a), but with $\dot{V}_1/\dot{V}_2 \neq 1$. The system is off-resonance, and the previously accumulated thermal energy rapidly leaves the system. \emph{Right:} same as left, but the system releases energy by damping the interaction and suppressing the oscillations.
\textbf{c)} Example of accumulated thermal energy $E_{\mathrm{th}}$ and normalized outlet output to input power $P_{\mathrm{out}}/P_{\mathrm{in}}$ versus time for a system initially at resonant (Accumulation phase) and then switched to off-resonance (Release phase).
}
  \label{fig:concept}
\end{figure}

Our prototypical heat oscillator is represented in Fig.~\ref{fig:concept}: a heated (e.g., through solar absorption) two-channel counter-current configuration, in which intra-channel advective heat transport coupled to inter-channel conductive exchange produces repeated alternation of heat transfer between the channels. Consistent with our previous work,\cite{Ye2023} we use the term ``heat oscillation'' to describe the cyclic exchange and recirculation of heat between the coupled counter-flowing channels. It does not denote a propagating thermal wave or second-sound-like transport.\cite{Huberman2019} Rather, it arises from the coupled advection--conduction dynamics and corresponds, in the dynamical-systems description, to a spiral-like transient in a reduced thermal state space. In this manuscript, ``heat oscillation'' and ``heat recirculation'' therefore refer to the same physical process: the former emphasizes its dynamical evolution, whereas the latter provides an intuitive description of the repeated internal heat exchange. Such a system behaves as a resonant heat oscillator when heat capacity rates are matched (i.e., equal volumetric flow rates $|\dot{V}_1| = |\dot{V}_2|$, in case of same fluid in the two channels), and losses are minimized by dynamically ``trapping'' thermal energy in the moving fluids. The degree of internal recirculation can be quantified by an effective thermal quality factor, equivalent to the $Q$ factor of any resonant cavity: a high $Q$ corresponds to a long thermal dwelling time and a large stored energy under fixed input power. This matched-rate condition distinguishes the oscillator from conventional heat-recovery loops, which return heat upstream at a fixed design point: here, internal exchange peaks sharply at matched flows (Fig.~S6a), vanishes in a cofluidic arrangement of the same channels (Supplemental Information, Section~S4.6), and only at the matched condition does the thermal dwelling time decouple from --- and far exceed --- the one-pass fluid dwelling time (Eq.~\ref{eq:tau_h} below). This sharply peaked response to a matching condition, together with the cavity-like quality factor of Eq.~\ref{eq:Qdef}, justifies the resonance terminology beyond analogy. In this configuration, large recirculating heat fluxes have already proven useful, for example, to significantly enhance the energy efficiency of thermal desalination processes.\cite{Alabastri2020b,Schmid2025}

A static, large $Q$ heat oscillator can accumulate thermal energy but cannot release it on demand. Active $Q$-switching resolves this by combining the high energy-storage density of a large $Q$ system with a controllable high-power output stage, upon high-to-low $Q$ switching. Similarly to any oscillator (LC circuit, optical cavity, etc.), the quality factor can be defined as:
\begin{equation}
  Q = \frac{\omega_h\, E_\mathrm{th}}{P_\mathrm{loss}}
  \label{eq:Qdef}
\end{equation}
where $E_\mathrm{th}$ is the accumulated thermal energy, $P_\mathrm{loss}$ is the total power lost, including contributions from both outlet losses and environmental heat dissipation, and $\omega_h$ is the rate at which the heat cycles in the system. Here, $\omega_h$ does not correspond to an externally imposed modulation frequency, but to the intrinsic heat-cycling rate along and across the channels: at resonance, heat advects along each channel in one fluid dwelling time, $\tau_{f,\mathrm{dwell}} \sim L/2u$, and transfers across the interface at the matched rate, so that $\omega_h = 1/\tau_{f,\mathrm{dwell}}$ (see the discussion following Eq.~\ref{eq:tau_h}). At steady state, $P_\mathrm{loss}$ is equal to the input power, $P_\mathrm{in}$, of the system (e.g., due to solar absorption), and a large $Q$ indicates that a large amount of energy is stored and recirculates fast within the system.

We now consider how the accumulated energy can be released in a controllable way. One strategy to reduce (switch) the $Q$ factor is to tune the cavity off-resonance.
This can be achieved by manipulating the mutual input velocities of the flows (Fig.~\ref{fig:concept}b, left) or by blocking fluid thermal interactions (Fig.~\ref{fig:concept}b, right).
Practically, such interfacial $Q$-switching requires a reversible thermal switch integrated between the channel walls. Candidate mechanisms include contact--separation mechanical and MEMS thermal switches, gap-opening actuators, and phase-change or field-responsive thermal interfaces,\cite{Wehmeyer2017,Ram2023,Kommandur2023} with demonstrated on/off conductance ratios of order $10^2$--$10^3$ and actuation times of milliseconds to seconds. Because the thermal dwelling times of the systems considered in this work range from $\sim$$10^2$--$10^3$~s (release) to $\sim$$10^3$--$10^4$~s (accumulation), the switch actuation is two to four orders of magnitude faster than the thermal dynamics: it acts effectively instantaneously and does not broaden the release pulse. In practice, contact resistance, parasitic conduction through structural supports, actuator heat generation, incomplete separation, and mechanical fatigue reduce the effective switching ratio; the $\sim$$10^3$ contrast considered here is therefore a design target achievable with current switch technology rather than a value demonstrated within the present fluidic device.
The switching energy itself is negligible: actuation is a single mechanical event per cycle, requiring of order mJ--J for mechanical or MEMS thermal switches, many orders of magnitude below the ${\sim}10$--$10^2$~kJ of thermal energy stored and released per cycle in the systems considered here.
In a thermal cavity system, resonant heat recirculation relies on two complementary processes: advective heat transport by fluid flow and conductive coupling between the countercurrent channels. On the one hand, detuning flows sets the system off-resonance because heat advection in the two channels is now mismatched, and heat cannot recirculate at a defined rate. On the other hand, by drastically reducing the conductive heat exchange between the fluids with an active thermal switch\cite{Wehmeyer2017} (e.g., by using mechanical actuation to open an insulating gap\cite{Ram2023} or altering the thermal conductivity of the conductive material\cite{Kommandur2023}), inter-channel thermal conduction is virtually interrupted and the two channels become independent. In both cases, the $Q$-factor is significantly reduced, and the stored thermal energy is rapidly discharged (Fig.~\ref{fig:concept}b, right).

Fig.~\ref{fig:concept}c illustrates the working principle of active $Q$-switching in quantitative terms using the interfacial-decoupling mechanism and plotting heat accumulation and release. Fig.~\ref{fig:concept}c shows the 2D-simulated (with the extension normal to the plane assumed as $d_z = \SI{1}{\meter}$) temporal dynamics of the thermal energy density for a pair of $w = \SI{1}{\milli\meter}$ thick and $L = \SI{1}{\meter}$ long countercurrent Galinstan-filled channels flowing with velocity $u = \SI{0.04}{\meter\per\second}$. The system is illuminated by $I_\mathrm{in} = \SI{0.5}{\kilo\watt\per\square\meter}$ absorbed at the top surface by a negligible-thickness photothermal absorber layer (e.g., carbon black nanoparticles\cite{Kommandur2023}). Environmental losses are lumped into $h_\mathrm{env} = \SI{1}{\watt\per\square\meter\per\kelvin}$. The channels are alternatively separated by a copper ($k_\mathrm{Cu} = \SI{400}{\watt\per\meter\per\kelvin}$), $t_\mathrm{cond} = \SI{100}{\micro\meter}$ thin wall or a $t_\mathrm{ins} = \SI{1}{\milli\meter}$ air gap ($k_\mathrm{air} = \SI{0.03}{\watt\per\meter\per\kelvin}$), ensuring either high conductivity or insulation during heat accumulation (high $Q$) and release (low $Q$) operation, respectively. In this case, it was possible to obtain an interaction heat transfer coefficient contrast of $h_\mathrm{int}|_\mathrm{high-Q}/h_\mathrm{int}|_\mathrm{low-Q} \approx 10^3 \gg 1$ (see Supplemental Information, Section~S1, for the analytical expression and derivation of $h_\mathrm{int}$, and for the analytical thermal energy density estimates in both regimes).

During the relatively long accumulation phase, because of the strong inter-channel interaction, $h_\mathrm{int} \approx \SI{1.64e4}{\watt\per\square\meter\per\kelvin}$, the two channels resonantly exchange heat. In this configuration, the system displays a relatively long characteristic "heat" dwelling time, $\tau_{h,\mathrm{dwell}}\approx \SI{500}{\second}$, compared to the fluid dwelling time $\tau_{f,\mathrm{dwell}} \sim \frac{L}{2u} \approx \SI{13}{\second}$ to reach steady state, as heat cycles several times before the internal temperature reaches a value where input power and losses equalize. In our previous work,\cite{Alabastri2020} we found that such recirculating heat flow is maximized when the fluid speed, $u^*$, in the two countercurrent channels is the same (for equal fluids in equal channels) and has a specific value, approximated by:
\begin{equation}
  u^* \approx \frac{\alpha}{\sqrt{2}}\,\frac{\sqrt{h_\mathrm{env}\, h_\mathrm{int}}\, L}{\rho\, c\, w}
  \label{eq:ustar}
\end{equation}
with $\alpha \cong 1/1.506$. The normalized recirculating heat (i.e., the integral of the heat flux through each channel section or, equivalently, across the interface between the channels and termed $A$ in our previous work\cite{Alabastri2020}), $P_\mathrm{circ}/P_\mathrm{in}$, effectively represents the $Q$-factor of our system:
\begin{equation}
  Q \equiv A \approx \beta\sqrt{\frac{h_\mathrm{int}}{2\,h_\mathrm{env}}}
  \label{eq:QfromA}
\end{equation}
with $\beta \cong 0.383$. Physically, $Q$ represents the number of cycles required for the heat to be lost to the outlets or the environment. When the system is at resonance, the ``heat'' dwelling time, $\tau_{h,\mathrm{dwell}}$, is disentangled from (and much larger than) the fluid dwell time, $\tau_{f,\mathrm{dwell}} \sim \frac{L}{2u^*}$:
\begin{equation}
  \tau_{h,\mathrm{dwell}} \approx Q\,\tau_{f,\mathrm{dwell}} = \gamma\,\frac{\rho\, c\, w}{h_\mathrm{env}}
  \label{eq:tau_h}
\end{equation}
with $\gamma = \beta/2\alpha$. For the case of Fig.~\ref{fig:concept}c we find $Q \approx 35$ and $\tau_{h,\mathrm{dwell}} \approx \SI{544}{\second}$, giving a thermal equilibration or accumulation time $\tau_\mathrm{acc} \sim 5\tau_{h,\mathrm{dwell}} \approx \SI{2700}{\second}$, consistent with the accumulated thermal energy density at saturation, $E_\mathrm{th} \sim 40~\mathrm{kWh/m^3}$, and in agreement with the simulation and analytical predictions (see Supplemental Information, Section~S1, for details). 

Equations~\ref{eq:ustar}--\ref{eq:tau_h} are reduced-order expressions with a defined range of validity. They assume two identical, slender channels ($L \gg w$) carrying the same incompressible fluid with constant thermophysical properties, laminar flow, matched heat-capacity rates (the resonant condition), spatially averaged coefficients $h_\mathrm{int}$ and $h_\mathrm{env}$, and advection-dominated axial transport ($\mathrm{Pe}_L \gg 1$); the dimensionless constants $\alpha$ and $\beta$ derive from our previous analysis of the resonant steady state.\cite{Alabastri2020} Within this regime, the expressions hold near the resonant optimum $u^*$ and for $h_\mathrm{int} \gg h_\mathrm{env}$ (equivalently $Q \gg 1$), and their predictions agree with the full conjugate heat-transfer simulations for the geometries and fluids considered in this work ($w = 1$--$2~\mathrm{mm}$, $L = 0.5$--$1~\mathrm{m}$, water and Galinstan), as exemplified by the case of Fig.~\ref{fig:concept}c ($Q \approx 35$, $\tau_{h,\mathrm{dwell}} \approx \SI{544}{\second}$, $E_\mathrm{th} \sim \SI{40}{\kilo\watt\hour\per\cubic\meter}$; Supplemental Information, Section~S1). Away from resonance, or for turbulent flow, strongly temperature-dependent properties, or nonuniform losses, they should be read as scaling relations, and the full numerical model of Methods applies instead. We also note that the resonant condition and $Q$ (Eqs.~\ref{eq:ustar}--\ref{eq:QfromA}) depend on $h_\mathrm{int}$, $h_\mathrm{env}$, and geometry but not on the input power: a fluctuating input therefore does not detune the high-$Q$ state, and the stored energy simply scales linearly with the time-averaged input, making the accumulation phase inherently tolerant to input variability, as verified numerically under sinusoidally modulated input in Supplemental Information, Section~S12.

In this framework, the equivalent oscillation frequency of the thermal cavity follows directly from Eq.~\ref{eq:Qdef} as $\omega_h = 1/\tau_{f,\mathrm{dwell}}$ and has a clear physical representation: along the channels, heat is transferred by advection on a timescale $\tau_{f,\mathrm{dwell}}$; when the system is at resonance, heat cycles around the system, and $\tau_{f,\mathrm{dwell}}$ matches the time for conductive transfer across the channels, identifying the thermal cavity frequency. Because these transfers occur in a loop, outlet losses are minimized, leading to an enhanced heat dwelling time, $\tau_{h,\mathrm{dwell}}$.

The key consequence, highlighted in Fig.~\ref{fig:concept}c and made visually explicit by the unified time axis, is the $\gg 1$ contrast between accumulation and release timescales. Switching Cu with air at the interface reduces $h_\mathrm{int}$ from $\SI{1.6e4}{}$ to about \SI{30}{\watt\per\square\meter\per\kelvin} and reduces $Q$ from 35 to 1.4. The accumulated thermal energy is released by advection at the outlets on the short fluid timescale: $\tau_\mathrm{rel} \sim \tau_{f,\mathrm{dwell}} \ll \tau_\mathrm{acc}$. The power that was circulating, $P_\mathrm{circ}$, at large $Q$ is now released as $P_\mathrm{out}$. While losses prevent $P_\mathrm{out}$ from reaching the upper bound $Q P_\mathrm{in}$, in Fig.~\ref{fig:concept}c (solid red line) we obtain $P_\mathrm{out}/P_\mathrm{in} \sim 30$. This order-of-magnitude amplification is the operational signature of active $Q$-switching: it is not available to any static architecture, because it depends on the \emph{ratio} between two operationally distinct dwell times set by $Q$ itself.

\subsection{$Q$-switching experimental validation}

\begin{figure}[htbp]
  \centering
   \includegraphics[width=0.8\textwidth]{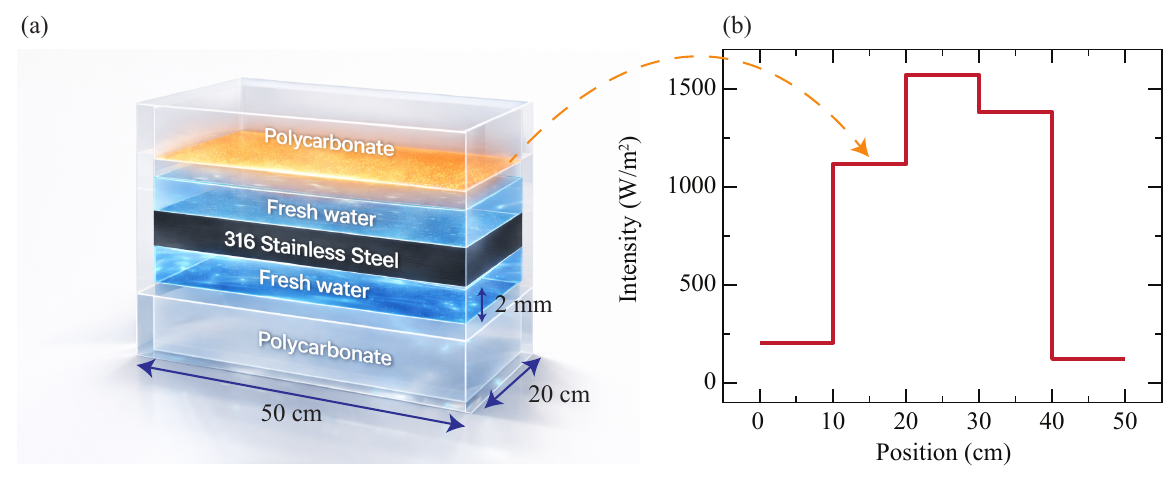}
  \caption{\textbf{Experimental configuration and illumination profile of the dual-channel microfluidic heat oscillator.}
  \textbf{(a)} Three-dimensional schematic of the fabricated device, consisting of a 316L stainless steel midplane sandwiched between two polycarbonate slabs, forming two counter-flow water channels. One side of the stainless-steel plate is coated with high-temperature black paint to enable efficient photothermal absorption. The system is illuminated from above by a high-power LED source positioned \SI{4}{\centi\meter} above the device.
  \textbf{(b)} Measured spatial irradiance distribution along the \SI{50}{\centi\meter} channel length produced by the LED lamp, showing a stepped intensity profile arising from the lamp geometry. The irradiance varies from approximately 120 to \SI{1570}{\watt\per\square\meter} across five distinct zones. This non-uniform illumination profile is incorporated into numerical simulations.}
  \label{fig:setup}
\end{figure}

As a proof of concept, we fabricated a two-channel microfluidic heat-oscillator device illuminated by LEDs, as shown in Fig.~\ref{fig:setup}(a). The heat oscillator consists of a \SI{20}{\centi\meter} (wide) $\times$ \SI{50}{\centi\meter} (long) custom-fabricated dual-layer heat oscillator system, composed of a 316L stainless steel midplane with a thickness of \SI{0.25}{in}, sealed between two polycarbonate slabs with a thickness of \SI{0.5}{in} and with \SI{2}{\milli\meter}-thick water channels. The structural integrity of the channel geometry was maintained using a bolted frame, with mechanical sealing verified via repeated leak tests. Water flow control was achieved using two variable-speed peristaltic pumps, one for lower flow rates and one for higher flow rates. One side of the stainless steel inside the heat oscillator system was coated with black paint to absorb light from the LED lamp, placed \SI{4}{\centi\meter} above the system. Such a configuration was easier to implement and it is possible for transparent fluids. In fact, we have previously shown that the heating location (top or between channels) does not significantly affect heat oscillations \cite{Ye2023}. The lamp's geometry (\SI{35.4}{\centi\meter} lens length, $28.72^\circ$ emission angle) created a stepped spatial irradiance distribution along the \SI{50}{\centi\meter} channel length, characterized independently using a thermal power sensor. The measured irradiance ranged from approximately 120 to \SI{1570}{\watt\per\square\meter} across five distinct zones, as shown in Fig.~\ref{fig:setup}(b), and this non-uniform profile was accounted for in our simulations. Water temperatures were constantly measured every \SI{5}{\second} at the input and output of both channels using wire probes directly connected to the thermocouple. Full device fabrication, illumination characterization, instrumentation, calibration procedures, and run-to-run reproducibility data are provided in Supplemental Information, Section~S2. Additional experimental detuning cases (resonant baseline plus positive and negative flow detuning under Light On $\to$ Light Off cycles) are reported in Supplemental Information, Section~S3, where we also discuss the asymmetry between positive and negative detuning and its origin in the internal temperature and heat-flux gradient maps.

\begin{figure}[htbp]
  \centering
   \includegraphics[width=0.8\textwidth]{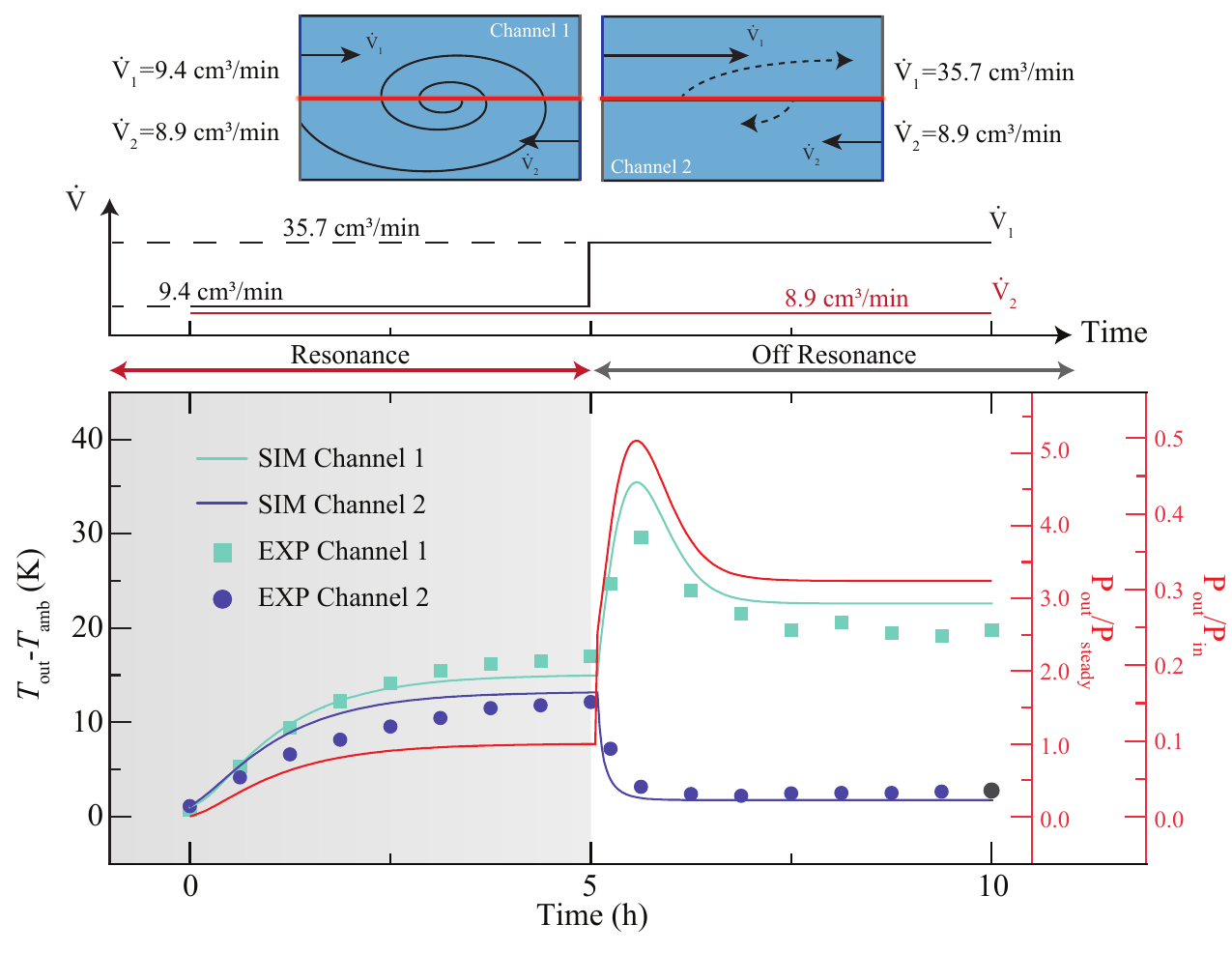}
  \caption{\textbf{Heat oscillator response to $Q$ switching by flow detuning (experiments and simulations).}
  \textit{Top:} scheme of the flow rate configurations for the experimental systems with matching flows with high $Q$ (left) and faster Channel 1 with low $Q$ (right). \textit{Middle:} Flow rate vs. time for varying Channel 1 (black) and fixed Channel 2 (red). \textit{Bottom:} Experimental (dots) and simulated (solid) temperature increase vs. time with respect to the ambient of Channel 1 (green) and Channel 2 (purple) on left y-axis. Calculated outlet power normalized to outlet power before the $Q$ switching, for Channel 1 vs. time (red) on the right y-axis. The quantities are plotted across the high $Q$  (resonance) to low $Q$ (off-resonance) configuration triggered by flow detuning. Experimental uncertainties (standard deviation, $n = 2$ repetitions) are typically smaller than the plotted symbols and are omitted for clarity; the same flow-detuning protocol with error bars is shown in Fig.~S8 (Supplemental Information, Section~S5.3).
  }
  \label{fig:Qswitch_exp}
\end{figure}

The simulated curves shown alongside the experimental data in this section are obtained with the two-dimensional, time-dependent conjugate heat-transfer model specified in Methods (geometry, boundary conditions, solver, and time stepping), with the measured irradiance profile of Fig.~\ref{fig:setup}b imposed as the heat-flux boundary condition and mesh/time-step convergence documented in Supplemental Information, Section~S7.
In our experimental setup, we demonstrated the $Q$-switching dynamics and accumulation/release operations under constant illumination by detuning the flow rates. Fig.~\ref{fig:Qswitch_exp} (Top) shows the scheme for a detuning in the top channel ($\dot{V}_1$ increased from \SI{9.4}{\centi\meter\cubed\per\minute} to \SI{35.7}{\centi\meter\cubed\per\minute}, $\dot{V}_2 = \SI{8.9}{\centi\meter\cubed\per\minute}$) from the resonant case (roughly equal $\dot{V}_1 = \SI{9.4}{\centi\meter\cubed\per\minute}$ and $\dot{V}_2 = \SI{8.9}{\centi\meter\cubed\per\minute}$). The flow rates vs. time for the two channels (black: top channel; red: bottom channel) are also shown. The experimental ('EXP' - dots) and simulated ('SIM' - solid) temperatures (blue and purple, left axis) and calculated output power dynamics (red, right axis) are shown in Fig.~\ref{fig:Qswitch_exp} (Bottom): after about $5~h$ of illumination at resonance (high-$Q$), thermal energy is accumulated and the outlet temperatures of both channels (1, top and 2, bottom) increased with respect to ambient by $\sim 15~ K$ reaching the steady state. At $t=5~h$, the flow rate of Channel 1 is quickly increased: the system is now off-resonance (low $Q$), and the stored enthalpy in both channels is released through the faster Channel 1. As a consequence, while the outlet temperature of Channel 1 quickly increases, the temperature of Channel 2 decreases because of thermal energy extraction through Channel 1. The 'pulse' shape of the Channel 1 outlet temperature increase reflects the previously accumulated thermal energy in both channels before $Q$ switching. While losses in our lab-scale setup are relatively large and the peak output power is lower than the constant input illumination, our experiment achieved a $\sim5$-fold increase in power from Channel 1 compared to the pre-switch baseline, $P_{steady}$, demonstrating controllable, significant transient output delivery. It should be noted that at longer times, the two channels stabilize at different outlet temperatures because the system is still at low $Q$, and Channel 1 (the faster channel) carries the vast majority of the absorbed heat out of the system. To restart the cycle, the system needs to be reset to a high $Q$ state, as shown later.  We note that the relatively small (less than an order of magnitude) pre-switch amplification observed here ($\sim 5\times$) is limited by the losses of our uninsulated lab-scale device, in which surface convection and radiation dominate the environmental loss term $h_\mathrm{env}$ and cap the achievable $Q$. 
Fitting the measured stationary outlet temperatures with the transient model yields an effective $h_\mathrm{env} \approx \SI{4.7}{\watt\per\square\meter\per\kelvin}$ for the uninsulated device, and this single value is used in all experimental-validation simulations (Supplemental Information, Section~S2.2).
However, crucially, the same transient model that quantitatively reproduces the experimental traces in Fig.~\ref{fig:Qswitch_exp} is used in the projections of the following sections with $h_\mathrm{env}$ reduced to values representative of an insulated, application-relevant system and the predicted higher amplifications follow directly from Eq.~\ref{eq:QfromA} ($Q \propto 1/\sqrt{h_\mathrm{env}}$).

Each experimental condition was repeated at least twice; the outlet-temperature trajectories and the extracted accumulation and release times varied by less than 5\% across repetitions, and the corresponding experimental uncertainties (standard deviation, $n = 2$, typically smaller than the plotted symbols) are reported in Supplemental Information, Section~S5.3 and Fig.~S8, for the same flow-detuning protocol as Fig.~\ref{fig:Qswitch_exp}. In the near-resonant case, the simulated outlet temperatures reproduce the measurements to within ${\sim}1$~K (Supplemental Information, Section~S3).

\subsection{$Q$-switching dynamics analysis}

Having validated the transient model against experiment, we now use it to analyze the second $Q$ switching mode: active interfacial thermal decoupling. We consider a water-based system (same as the experiments) with a device length of \SI{50}{\centi\meter}, depth of \SI{20}{\centi\meter}, and a fluid-channel width of \SI{2}{\milli\meter}, operating under continuous solar illumination at \SI{0.1}{\kilo\watt\per\square\meter} and low environmental losses ($h_\mathrm{env} = \SI{1}{\watt\per\square\meter\per\kelvin}$; effects of larger losses are treated in Supplemental Information, Section~S8). The system starts accumulating thermal energy in its high $Q$ regime and is switched to a low $Q$ by deliberately suppressing the interfacial heat transfer coefficient at $t=6000 ~s$. To trigger discharge via thermal decoupling, we model the interface using a surrogate layer that strongly suppresses heat exchange ($k_\mathrm{int} = \SI{e-5}{\watt\per\meter\per\kelvin}$ at thickness $t_\mathrm{int} = \SI{0.1}{\milli\meter}$). Such a small conductivity value emulates a near-ideal insulator such as a silica-aerogel barrier ($t_\mathrm{int} = \SI{5}{\milli\meter}$, $k_\mathrm{int} = \SI{0.01}{\watt\per\meter\per\kelvin}$) that yields an indistinguishable thermal response, as shown in Supplemental Information, Section~S6.

\begin{figure}[htbp]
  \centering
   \includegraphics[width=0.8\textwidth]{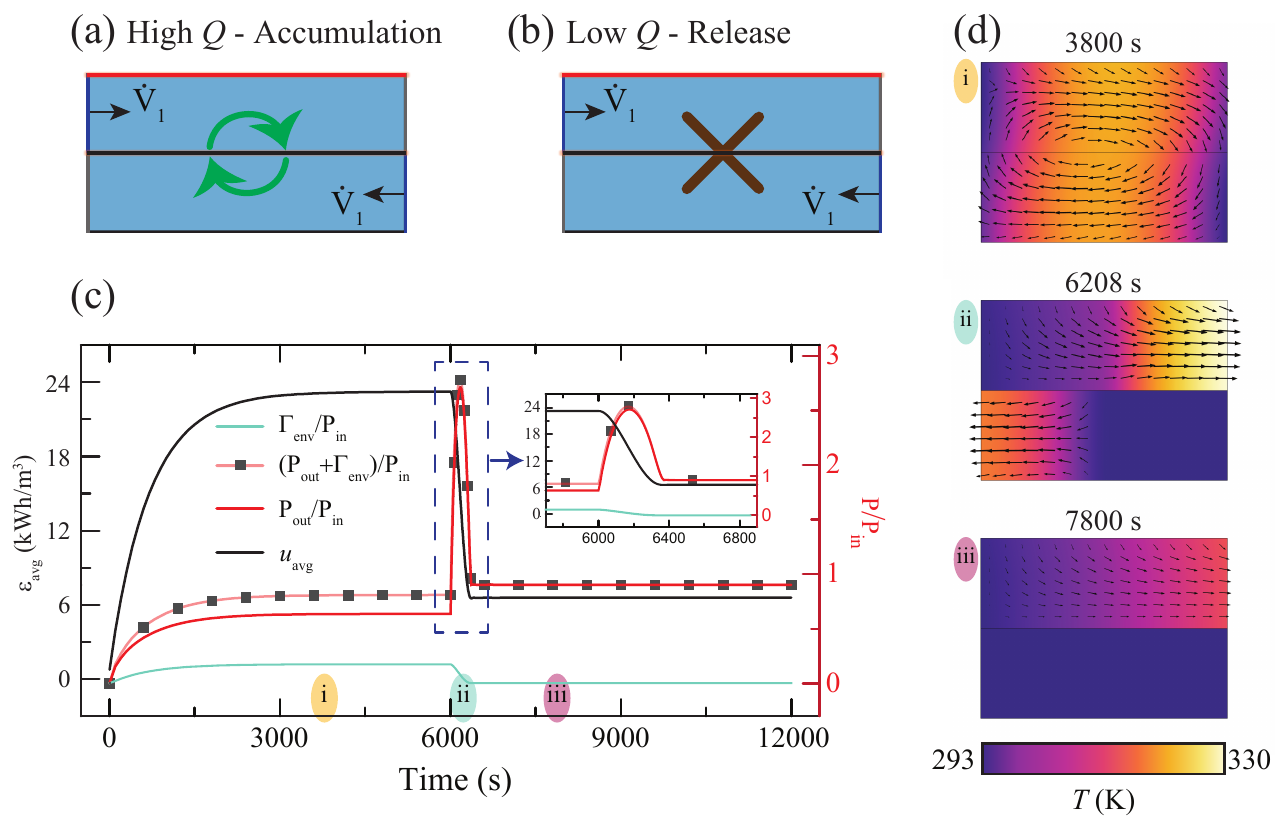}
  \caption{\textbf{Resonant system response to interchannel thermal interaction modulation (simulations).} Water-based system; $L = \SI{0.5}{\meter}$, $\dot{V}_1 = \SI{32}{\centi\meter\cubed\per\minute}$, illumination \SI{0.1}{\kilo\watt\per\square\meter}, $h_\mathrm{env} = \SI{1}{\watt\per\square\meter\per\kelvin}$. \textbf{(a)} High-$Q$ accumulation: resonant coupling sustains heat recirculation (green loop). \textbf{(b)} Low-$Q$ release: an insulating interfacial layer (red ``X''; $t = \SI{0.1}{\milli\meter}$, $k = \SI{e-5}{\watt\per\meter\per\kelvin}$) suppresses the coupling. \textbf{(c)} Normalized powers $P_\mathrm{out}/P_\mathrm{in}$, $\Gamma_\mathrm{env}/P_\mathrm{in}$, and $(P_\mathrm{out}+\Gamma_\mathrm{env})/P_\mathrm{in}$, and average energy density $\varepsilon_\mathrm{avg}$, vs.\ time; switching at $t = \SI{6000}{\second}$. \textbf{(d)} Temperature maps and heat-flux vectors at \SI{3800}, \SI{6208}, and \SI{7800}{\second}. Flow rates are constant throughout; only the interfacial coupling is switched.}
  \label{fig:dynamics}
\end{figure}

Figure~\ref{fig:dynamics}c shows the power output and energy evolution upon $Q$ switching. Up to $t = \SI{6000}{\second}$, the system accumulates energy (resonant, large $Q$). At $t = \SI{6000}{\second}$, we switch to an insulating interface (off-resonant, small $Q$) by assigning a low interfacial conductivity. Immediately, the outlet power ($P_\mathrm{out}$) surges as the system releases its stored energy. The characteristic time of the system's thermal response changes by an order of magnitude: during the accumulation stage, the time constant was $\tau_{h,\mathrm{dwell}} \approx \SI{3800}{\second}$, but upon switching it reduces to $\tau_{f,\mathrm{dwell}} \approx \SI{375}{\second}$. As a result, all the accumulated heat is expelled through the outlet in a few hundred seconds --- the thermal analogue of the optical pulse released when a laser cavity is switched from high to low loss.

Figure~\ref{fig:dynamics}d provides snapshots of the temperature field and heat flux vectors at three relevant times: (i) just before switching, a strong recirculating heat flux loops around the channels; (ii) shortly after switching, the heat flux streams out towards the outlet (red arrow) as the channels decouple; and (iii) after full energy release the system cools down with minimal remaining flux. This strategy leverages the large contrast between the $Q$ factors, enabling efficient energy storage and high-power release. 
Consequently, the system briefly produces a pulsed heat output exceeding the steady input power. This behavior follows directly from the instantaneous energy balance
\begin{equation}
P_{\mathrm{out}}
=
P_{\mathrm{in}}
-
\Gamma_{\mathrm{env}}
-
\frac{\mathrm{d}E_{\mathrm{th}}}{\mathrm{d}t}.
\label{eq:energy_conservation}
\end{equation}
During discharge, $\mathrm{d}E_{\mathrm{th}}/\mathrm{d}t<0$, such that the release of previously stored thermal energy can temporarily produce $P_{\mathrm{out}}>P_{\mathrm{in}}$. Upon time integration, the total discharged energy remains equal to the supplied energy plus the reduction in stored thermal energy, minus environmental losses; therefore, energy conservation is strictly satisfied.
While the focus of this work is peak-power amplification, the energy efficiency of the $Q$-switching cycle can be characterized by two complementary metrics: for the cycle of Fig.~\ref{fig:dynamics}c, the energy delivered in the release burst corresponds to approximately 8\% of the cumulative energy input over the full cycle, and to 67\% of the thermal energy stored in the oscillator up to the switching time. The balance of the input is delivered continuously at the outlets during accumulation, environmental dissipation ($\Gamma_\mathrm{env}$) being the only true loss of the cycle, governed by $h_\mathrm{env}$ alone.
For the water-based case of Fig.~\ref{fig:dynamics}c we observe a peak $P_\mathrm{out}$ of about $2$--$3P_\mathrm{in}$, after which it decays. Replacing water with a liquid metal allows this amplification to increase substantially because the fluid thermal conductivity supports a larger $u^*$. Details of the numerical implementation for such a system, including mesh-independence and time-step convergence tests, are provided in Supplemental Information, Section~S7.

It should be noted that during the high $Q$ phase, the energy density $E_{th}$ increases monotonically and saturates at $E_{th} \approx \SI{40}{\kilo\watt\hour\per\cubic\meter}$ by $t \approx \SI{6e3}{\second}$ (Fig.~\ref{fig:dynamics}c). This is comparable to the sensible-heat capacity of water for a \SI{40}{\kelvin} rise ($\rho c_p \Delta T \approx \SI{35}{\kilo\watt\hour\per\cubic\meter}$) and approaches the latent-heat capacities of common paraffin PCMs (${\sim}40$--$\SI{70}{\kilo\watt\hour\per\cubic\meter}$). We emphasize, however, that energy density per se is not the distinguishing figure of merit of this system. Conventional sensible and latent media with comparable or superior $E_{th}$ exist; the difference is that those media cannot release their stored energy as a short, high-power burst without an external power-conversion stage. Active $Q$-switching decouples energy density from deliverable peak power, a degree of freedom that static storage does not possess. Further analyses of how $\tau$ depends on channel geometry and environmental losses are provided in Supplemental Information, Section~S8.

\subsection{$Q$-switching strategy comparison}

We now compare the two $Q$-switching actuation modes --- flow detuning and interfacial heat exchange modification --- and address how they can be chained across successive accumulation/release cycles. In the previous section, the system transitioned from energy accumulation to release by suppressing the interfacial thermal coupling. This conductivity gating effectively decouples the two fluidic channels, allowing the stored energy to be expelled through the outlets. While effective, such a mechanism may not always be practical in implementations where material constraints or actuation complexity limit its use. However, as shown, similar power-extraction behavior can be achieved solely through velocity modulation. By dynamically increasing the inlet velocity $u(t)$ while keeping the interfacial conductivity fixed, the system enters a high-P\'{e}clet number regime in which the advective timescale becomes much shorter than the interfacial conduction time. Under this condition, cross-channel thermal exchange is suppressed during the release time window, and the system effectively enters a low $Q$ state. This complementary mechanism enables tunable thermal discharge via fluidic control, without altering material interfaces or requiring physical switches.

\begin{figure}[htbp]
  \centering
   \includegraphics[width=0.8\textwidth]{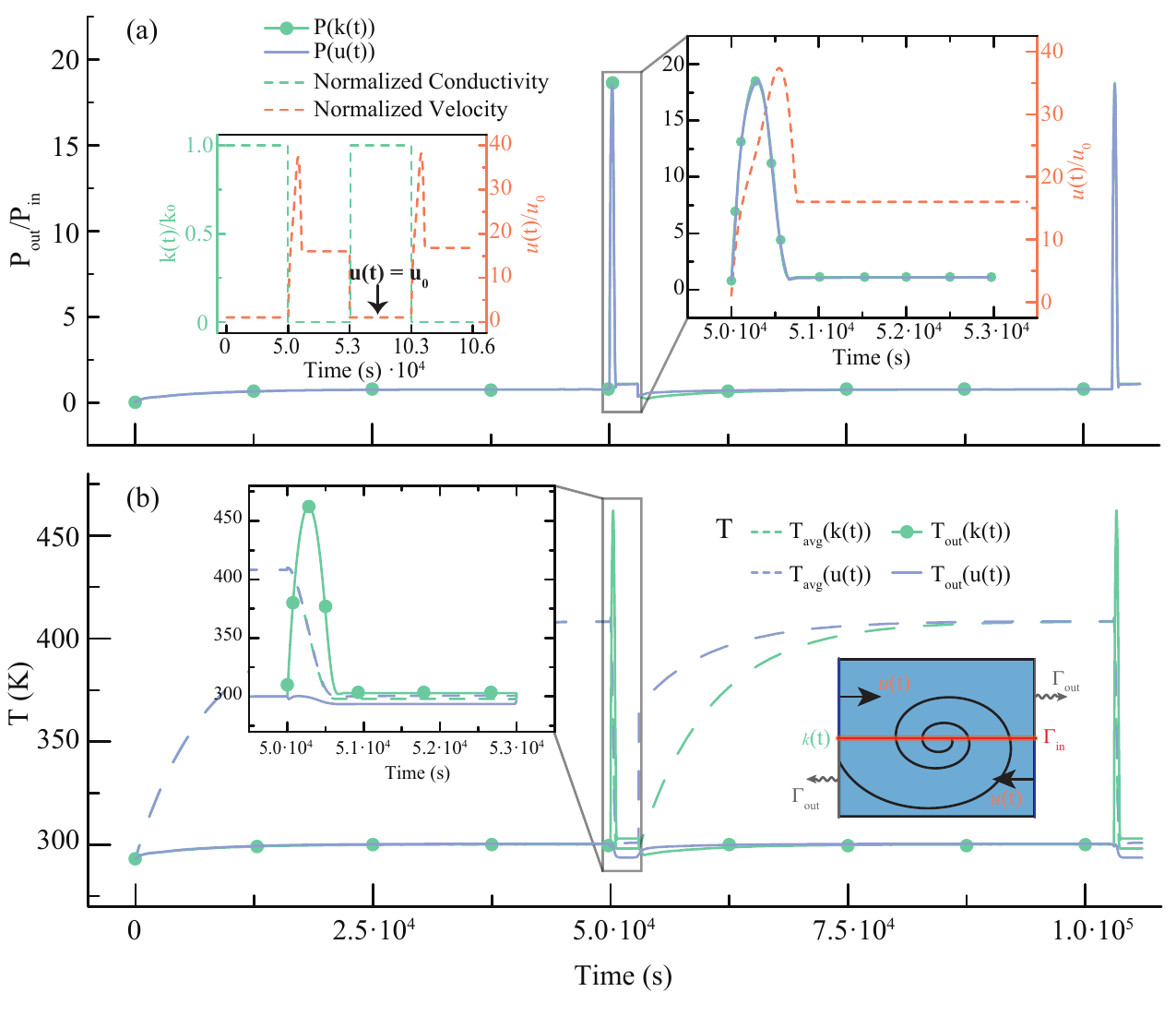}
  \caption{\textbf{Comparison between flow-detuning and interaction modulation Q-switching approaches (simulations).} (Galinstan; $L = \SI{1}{\meter}$, $w = \SI{1}{\milli\meter}$; illumination \SI{1}{\kilo\watt\per\square\meter}).
  \textbf{(a)} Normalized outlet power $P_\mathrm{out}/P_\mathrm{in}$ during accumulation and subsequent release triggered by two control protocols: modulation of the interfacial conductivity $k(t)/k_0$ (green) and modulation of the flow velocity $u(t)/u_0$ (blue). Left inset: imposed control signals for $k(t)/k_0$ and $u(t)/u_0$. Right inset: zoom of the first switching event, showing that the imposed velocity program (orange, right axis) produces the same extraction spike and relaxation behavior as conductivity gating.
  \textbf{(b)} Outlet temperature $T_\mathrm{out}$ (solid lines) and system-averaged temperature $T_\mathrm{avg}$ (dashed lines), where $T_\mathrm{avg}$ is obtained by averaging over the full channel cross-section and along the channel length. Left inset: zoom highlighting the narrow $T_\mathrm{out}$ burst and the subsequent monotonic relaxation of $T_\mathrm{avg}$. Right inset: schematic illustrating the two independent control parameters, $u(t)$ and $k(t)$. The black curve describes the spiral-like trajectory of a point in phase space, evolving under the heat flux vector field. A second actuation at $t \approx \SI{e5}{\second}$ demonstrates repeatability. During accumulation, the resonant flow condition is held fixed; only $k(t)$ or $u(t)$ is modulated at the switching event.}
  \label{fig:strategy}
\end{figure}

To compare and show the system-level equivalence of the two approaches, we implemented a Galinstan-filled two-channel system (\SI{1}{\meter} long, \SI{1}{\milli\meter} thick) with a \SI{1}{\kilo\watt\per\square\meter} heat source placed at the interface between the channels. For non-transparent fluids, such as Galinstan, the absorbing layer should be placed at the top, sun-facing surface. However, a symmetrical configuration simplifies the analysis, and we have placed the heating layer between the channels without affecting the results' significance, as shown in our previous work \cite{Ye2023}. Galinstan is selected for its high thermal conductivity (${\sim}\SI{16.5}{\watt\per\meter\per\kelvin}$), high density, and low viscosity, making it suitable for compact thermal storage and efficient convective transport. 
Galinstan is non-toxic and non-flammable --- in contrast to alkali-metal coolants such as Na or NaK --- and is liquid well below room temperature, and gallium-based liquid metals are already deployed as working fluids in high-heat-flux thermal management.\cite{Miner2004} Their principal engineering constraints, the formation of a thin self-limiting surface oxide and incompatibility with aluminum, are well characterized and are managed in practice by operating in closed loops with compatible wetted materials such as stainless steel or nickel-coated surfaces.\cite{Deng2009} We emphasize that the property central to the present approach is the thermal conductivity, which sets $u^*$ and hence the achievable release rate (Eq.~\ref{eq:ustar}); the framework itself is fluid-agnostic, and the working fluid can be selected according to the temperature window and compatibility requirements of a specific application.
As previously shown, higher conductivity allows for a larger $u^*$ and, therefore, a faster heat release. A brief velocity pulse applied from the resonant baseline leads to a rapid reduction in the advective dwell time, $\tau_{f,\mathrm{dwell}}$, triggering the release of stored energy. Figure~\ref{fig:strategy} compares these two approaches. In the top panel, the normalized outlet power $P_\mathrm{out}/P_\mathrm{in}$ shows nearly identical peak amplitude and decay behavior for both conductivity gating and velocity pulsing. The inset highlights the applied $u(t)$ waveform: a velocity pulse ramping from $u_0 = \SI{2}{\milli\meter\per\second}$ to $37.5\, u_0$ over \SI{475}{\second}, effectively reducing the fluidic dwell time by an order of magnitude and triggering an increased output power. In the bottom panel, the outlet temperature $T_\mathrm{out}$ and average system temperature $T_\mathrm{avg}$ follow the same temporal trajectory, confirming that both control modes yield equivalent dynamics. A second actuation at $t \approx \SI{e5}{\second}$ confirms that both approaches are repeatable within the deterministic numerical model. For the single-channel pair Galinstan oscillator, peak amplification reaches $P_\mathrm{out}/P_\mathrm{in} \approx 20$ --- significantly higher than the water case --- directly reflecting the larger $Q$ set by the thermal-conductivity contrast in Eq.~\ref{eq:QfromA}.

Compared to conductivity gating, velocity modulation may offer additional flexibility for real-time control. It requires no physical alteration of the system and can be reprogrammed dynamically to deliver user-defined power profiles using the same architecture. Moreover, it decouples the roles of storage and release: the interfacial conductivity $k$ determines how effectively energy is retained, while the flow velocity $u(t)$ dictates how quickly and strongly that energy is released. A visual decision map for selecting between the two strategies, together with a quantitative comparison of the conductivity- and velocity-tuning extraction limits, is provided in Supplemental Information, Section~S9. Together, these two control modes allow independent tuning of energy accumulation and discharge characteristics. However, each strategy has practical limits. Conductivity suppression provides a maximum extraction ceiling once the interface is fully decoupled. In contrast, velocity-driven release is not intrinsically capped but is limited by fluid-mechanical constraints, such as pressure drop and Reynolds number. To preserve laminar operation in Galinstan-filled channels, the Reynolds number must remain below a critical threshold (typically $10^3$). Excessively high velocities may also amplify convective losses to the environment, thereby reducing the system's effective figure of merit.
Environmental losses are quantified directly by the previous analytical framework: $Q \propto 1/\sqrt{h_\mathrm{env}}$ while the resonant velocity $u^* \propto \sqrt{h_\mathrm{env}}$ (Eqs.~\ref{eq:ustar}--\ref{eq:QfromA}), so increasing losses simultaneously lowers the achievable quality factor and raises the flow rate required to remain at resonance. Maintaining $Q \gtrsim 10$ therefore requires $h_\mathrm{env} \approx 1$--$10~\mathrm{W\,m^{-2}\,K^{-1}}$, attainable with standard insulation; for substantially larger losses, $Q$ approaches unity and the resonant flow rate becomes impractically large (scaling with $h_\mathrm{env}$ in Supplemental Information, Section~S8).

\subsection{Ragone Analysis: dynamic trajectories vs.\ static operating points}

\begin{figure}[htbp]
  \centering
   \includegraphics[width=0.9\textwidth]{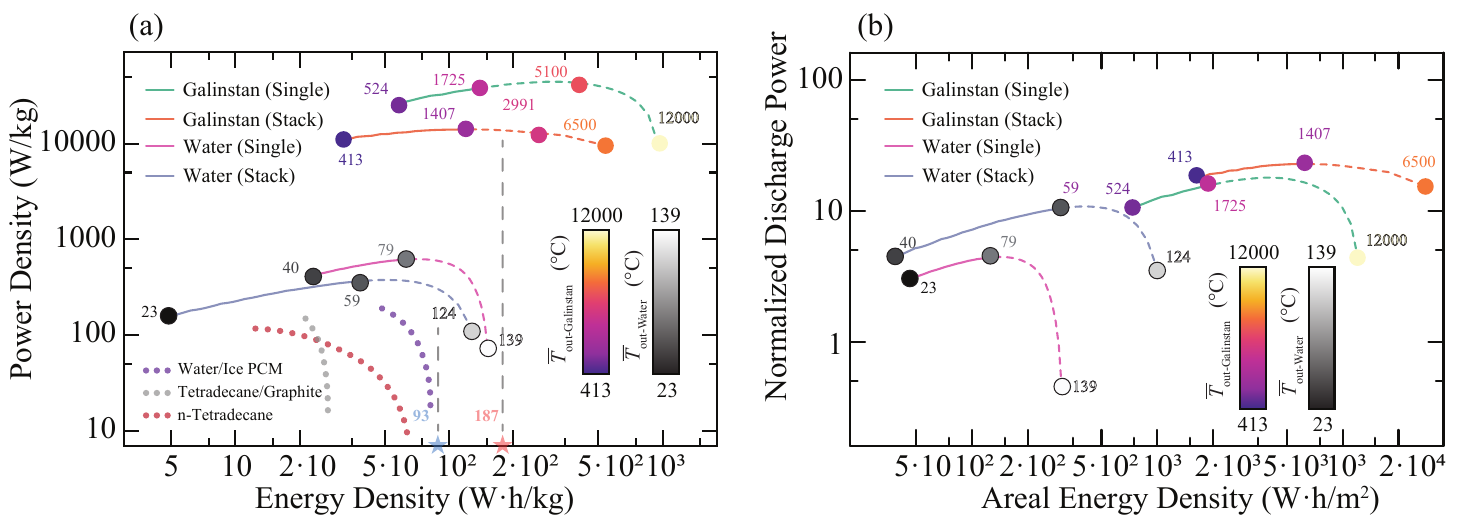}
  \caption{\textbf{Ragone-type operating maps: dynamic trajectories of $Q$-switched heat oscillators vs.\ static operating points of conventional TES (simulations and literature data).} \textbf{(a)} Mass-normalized power density vs.\ mass-normalized energy density. \textbf{(b)} Normalized discharge power (average extracted power during release, normalized by the incident input power) vs.\ area-normalized energy density, highlighting performance scaling with device footprint. Colored curves: simulated trajectories for water- and Galinstan-based single (2-CH) and vertically stacked (8-CH) oscillators; each point corresponds to a full accumulation--release cycle, with stored energy and released power defined in Eqs.~\ref{eq:deltaE}--\ref{eq:E90}. Marker color and the adjacent numeric label indicate the average outlet temperature during discharge (in $^\circ$C; separate color scales for water and Galinstan). Solid (dashed) curve segments: system temperatures below (exceeding) the boiling point of the working fluid. Dotted curves: rate-capability references for representative PCM storage media (water/ice, tetradecane/graphite composite, $n$-tetradecane).\cite{Woods2021} Vertical dashed lines (stars): sensible energy density required to heat water and Galinstan to their boiling points, excluding latent heat.}
  \label{fig:ragone}
\end{figure}

The single most distinctive feature of an actively $Q$-switched heat oscillator, and the one that separates it conceptually from all passive TES media, is that its operating point in the energy--power plane is not fixed: by modulating flow rates and interfacial coupling, the same equipment can trace continuous trajectories through the Ragone space. Conventional sensible, latent, and thermochemical TES media occupy \emph{points} in this plane, set by their material properties and heat-exchanger design. Figure~\ref{fig:ragone} quantifies this distinction by placing the RET oscillator alongside reference TES technologies on a common Ragone representation.

Figure~\ref{fig:ragone} presents Ragone-type plots using two complementary normalizations and two working fluids under different illumination conditions: \SI{0.28}{\kilo\watt\per\square\meter} for water and \SI{32}{\kilo\watt\per\square\meter} for Galinstan (representing low-intensity and solar-concentration conditions, respectively). Panel~(a) shows the mass-normalized power density as a function of mass-normalized energy density; panel~(b) reports the discharge power normalized by the incident input power as a function of area-normalized energy density, relevant for applications in which the device footprint is fixed. The vertical dashed lines indicate thermodynamic reference limits for uniform sensible heating of the working fluid to its boiling point; dashed portions of the colored curves correspond to operating conditions in which localized regions of the system exceed the boiling temperature. \emph{Dotted curves} show rate-capability references for representative PCM storage media (water/ice, tetradecane/graphite composite, $n$-tetradecane), compiled from the literature.\cite{Woods2021} These envelopes are fixed by material properties --- the deliverable energy falls monotonically as the discharge power rises --- whereas the oscillator trajectories are swept dynamically by the control parameters of the same hardware.

Each point along the colored curves corresponds to a complete accumulation--release cycle in which the system is switched from high $Q$ to low $Q$ upon saturating thermal energy storage. Unlike TES media, the operating point of the heat oscillator is dynamically tunable through flow control and oscillator coupling, enabling the system to \emph{sweep} through a continuous range of states within Ragone space. The stored energy is given by
\begin{equation}
  \Delta E_\mathrm{th} = E_\mathrm{th}^\mathrm{high-Q} - E_\mathrm{th}^\mathrm{low-Q},
  \label{eq:deltaE}
\end{equation}
where $E_\mathrm{th}^\mathrm{high-Q}$ is the sensible thermal energy accumulated at the end of the accumulation phase, and $E_\mathrm{th}^\mathrm{low-Q}$ is the residual sensible thermal energy remaining after the system has fully relaxed following the $Q$ switching. For the Ragone representation, the discharged energy is
\begin{equation}
  E_{90} = 0.9\,\Delta E_\mathrm{th},
  \label{eq:E90}
\end{equation}
the release of 90\% of the stored sensible thermal energy. The power density is the average extracted thermal power over the time interval required to release this 90\% fraction, so that the Ragone plots capture sustained energy release rather than the instantaneous transient spikes immediately following the $Q$ switching.

The plots include both 2-channel (2-CH) and 8-channel (8-CH) heat oscillators, with the 8-CH configuration representing a vertical stack of 4 2-CH systems. It was previously demonstrated \cite{Ye2023} that stacking multiple oscillators increases the total heat recirculation by reducing environmental losses. The dependence of stored energy and $Q$ on the number of stacked layers, applied here to the transient $Q$-switching regime, is reported in Supplemental Information, Section~S10.

Three features of Fig.~\ref{fig:ragone} should be emphasized. First, in absolute energy-density terms, the water-based oscillator is comparable to the PCM reference media --- as expected from the modest $\Delta T$ of the present operating conditions. Second, the oscillator nonetheless reaches \emph{power} densities substantially above the PCM rate-capability envelopes at matched energy density, occupying regions of the Ragone plane that no single static medium accesses with a single operating configuration. Third, in the mass-normalized representation (Fig.~\ref{fig:ragone}a), vertically stacked systems (8-CH) exhibit energy densities and powers comparable to or slightly lower than single-pair systems (2-CH); however, the area-normalized representation (Fig.~\ref{fig:ragone}b) shows that stacking significantly increases both energy and power \emph{per unit area}, because vertical integration increases active thermal volume and recirculating capacity without increasing exposed surface area. Together with the dynamic tunability highlighted above, the stacked Galinstan oscillator reaches $P_\mathrm{out}/P_\mathrm{in} \approx 40$ (numerical projection), the highest amplification obtained in this work.

From an energy-storage perspective, higher energy and power densities are desirable, and the lower-left region of the Ragone plot is typically unfavorable. However, there is no single optimal operating point; the appropriate regime depends on discharge duration and application constraints. The defining advantage of an actively $Q$-switched heat oscillator, unavailable to static TES media, is the ability to \emph{dynamically traverse Ragone space} to target the operating condition required at that moment. Resonant operation maximizes stored sensible energy by raising $Q$; high-to-low $Q$ switching enables large power density outputs. An extended Ragone analysis corresponding to explicit Light On $\to$ Light Off operation under fixed thermal coupling, which isolates the role of active $Q$-switching from geometry and steady illumination alone, is provided in Supplemental Information, Section~S11. A direct quantitative comparison of active $Q$-switching against the principal alternative methods for generating thermal pulses from continuous input (triggered-release supercooled PCMs, oscillating thermal switches, industrial peak-shaving systems, and PCM thermal buffers \cite{Krishnan2005PCM,Tianyu2021ThermalSwitch,Shamberger2016}) is provided as Table~S1 in Supplemental Information, Section~S13.

\section*{CONCLUSION}

The central result of this work is that resonantly coupled heat oscillators admit an actively switchable effective thermal quality factor $Q$, and that modulating $Q$ on sub-dwell-time timescales constitutes a distinct mode of thermal power management that separates energy harvesting from energy delivery as independent control axes. In the high-$Q$ state, resonant heat recirculation suppresses outlet losses and enables energy accumulation in moving fluids under steady thermal input. On demand, the system switches to a low-$Q$ state, either by detuning the countercurrent flow rates or by suppressing interfacial exchange, releasing stored enthalpy on the shorter advective timescale and producing transient outlet power that exceeds the steady input. Experiments on a water-based device validate the transient model, confirming the predicted $Q$ contrast between accumulation and release stages. The validated model predicts that liquid-metal implementations can achieve ${\sim}20\times$ peak amplification, and that vertically coupled oscillators can extend the achievable trajectory in the Ragone plane to ${\sim}40\times$ amplification.

The physical significance of these results is that a $Q$ switching heat oscillator occupies a different object class in the Ragone plane. Conventional sensible, latent, and thermochemical media correspond to fixed operating points set by material and hardware choices. An actively $Q$-switched heat oscillator traces continuous trajectories through the same plane, accessing high-peak-power regions that static architectures cannot reach with a single configuration. Within the broader context of active and non-equilibrium thermal transport,\cite{Li2019,Li2021,Qi2022,Cao2024,Wang2026}, this establishes $Q$-switching as a control method for sensible-heat systems that is conceptually distinct from conventional TES systems.

Translating these results into deployed systems will require engineering development beyond the present proof of concept: integration of a reversible interfacial switch with the fluidic channels, system-scale insulation maintaining $h_\mathrm{env} \approx 1$--$10~\mathrm{W\,m^{-2}\,K^{-1}}$, manifolding and flow balancing of stacked oscillators, and, for liquid-metal implementations, chemically compatible wetted materials and leak-tight containment.
Future work will focus on developing actuation mechanisms for reversible interfacial gating with large switch ratios, closed-loop control of $Q$ under variable environments, systematic sensitivity analysis of the release profile to perturbations in flow rates and operating conditions, and experimental demonstration of multi-cycle operation and stacked architectures. Immediate application targets include transient photothermal loads, microfluidic thermal management, and waste-heat recovery scenarios in which the ability to shape output power on demand, rather than raw stored energy, is the binding constraint.

\newpage

\section*{METHODS}


\subsection*{Numerical simulations}

All simulations were performed in COMSOL Multiphysics using a coupled, time-dependent conjugate heat transfer framework. Fluid domains were modeled with the laminar flow (incompressible) and heat transfer in fluids interfaces, while the midplane and channel walls were modeled with heat transfer in solids. Inter-channel heat exchange occurred via conduction across the shared wall. Environmental losses were imposed as a convective boundary condition on the external walls with coefficient $h_{\mathrm{env}}$ (default values stated in the corresponding figure captions). For the experimental validation in Fig. 4, the spatially varying irradiance profile measured along the channel length (Fig. 3b) was applied directly as a heat-flux boundary condition on the heated surface. For all other simulations, a uniform incident heat flux was applied as specified in the corresponding figure captions. No conduction term is removed from the numerical model; the axial-conduction approximation is used only in the reduced analytical derivation of Supplemental Information, Section~S4.

Unless otherwise noted, material properties (density, specific heat, thermal conductivity, viscosity) were taken as constant and evaluated at room temperature. The inlet temperature was fixed at $T_{\mathrm{amb}}$, and outlets were implemented as convective outflow boundaries. Time integration used COMSOL's generalized-$\alpha$ scheme with intermediate damping. To balance accuracy and runtime, the adaptive time-step controller was capped at $\Delta t_{\max} = 50$~s during slow segments (e.g., resonant accumulation) and $\Delta t_{\max} = 5$~s during fast transients (e.g., interfacial decoupling or abrupt flow detuning). The production mesh used $N_x = 500$ elements along the channel length, for which the outlet temperature differs by less than 0.1\% from an $N_x = 2000$ reference solution, and time-step convergence was verified for ceilings down to $\Delta t_{\max} = 25$~s; full convergence studies are reported in Supplemental Information, Section~S7.

Two Q-switching actuation modes were implemented. \textit{Flow detuning} was applied by abruptly changing one inlet flow rate at a prescribed switching time $t_d$ while keeping the geometry, heating profile, interfacial properties, and all boundary conditions fixed. \textit{Interfacial decoupling} was applied by switching the interfacial layer from a high-conductivity state ($k_{\mathrm{int}} = 400$~W~m$^{-1}$~K$^{-1}$, $t_{\mathrm{int}} = 100$~$\mu$m, representative of copper) to a low-conductivity state ($k_{\mathrm{int}} = 10^{-5}$~W~m$^{-1}$~K$^{-1}$, $t_{\mathrm{int}} = 0.1$~mm) at the prescribed switching time. The equivalence of this idealized insulating layer with a finite-thickness practical insulator (e.g., silica aerogel, $t_{\mathrm{int}} = 5$~mm, $k_{\mathrm{int}} = 0.01$~W~m$^{-1}$~K$^{-1}$) is demonstrated in Supplemental Information, Section S6.

\subsection*{Device fabrication}

A custom-fabricated dual-channel heat oscillator with lateral dimensions of 20~cm $\times$ 50~cm was used for $Q$-switching validation. The experimental module consisted of two polycarbonate plates of 0.5~in thick with a 0.25-in-thick 316L stainless-steel plate in the middle.  Two 2-mm-thick counter-flow water channels were formed between each polycarbonate plate and the stainless-steel plate. The device’s plates were maintained in place by a bolted frame. One surface of the internal stainless-steel plate was coated with high-temperature black paint (AERVOE Zynolyte Hi-Temp) to enhance optical absorption.

\subsection*{Illumination and characterization}

The experimental device was illuminated with a high-power LED lamp (FAISHILAN, 200~W, IP66). The lamp was placed at 4~cm above the device, had a lens of 35.4~cm long, and formed an emission angle of 28.72$^\circ$ with respect to the experimental module, which created a non-uniform incident irradiance profile along the 50~cm channel of the device. The irradiance profile was characterized using a thermal power sensor (THORLABS S350C),

\subsection*{Flow control and temperature measurement}

Two variable-speed peristaltic pumps were used to control the water flow rate. Two different pumps were used, depending on the flow regime: for low flow rates (Thermo Fisher Scientific, 0.4--85~mL~min$^{-1}$, Cat. No. 13-876-2) and for high flow rates (Anko VSL-600, 31--621~mL~min$^{-1}$). Each flow rate was calibrated by weighing the water output every 5~s over a 5~min interval using a precision balance (TORBAL, 3000~g maximum capacity, 0.01~g resolution).
Temperatures were measured using Type-K wire probes (REED Instruments TP-01, $-40$ to 250$^\circ$C) connected to a thermocouple thermometer (EXTECH SDL200). Probes were located at the inputs and outputs of both channels, and temperature was recorded every 5~s for each experiment.

\subsection*{Experimental protocol and reproducibility}

The light on/off cycle and the change for the inlet flow rates $(\dot{V}_1, \dot{V}_2)$ were manually synchronized for each experiment. Each experiment was repeated at least twice. The extracted accumulation and release times ($\tau_c$, $\tau_d$), outlet temperatures, and full thermal response curves exhibited less than 5\% variation across repetitions. Run-to-run reproducibility data, additional flow-detuning cases, and the corresponding uncertainty analysis are reported in Supplemental Information, Sections S2 and S3.

\newpage


\section*{RESOURCE AVAILABILITY}


\subsection*{Lead contact}


Requests for further information and resources should be directed to and will be fulfilled by the lead contact, Alessandro Alabastri (alabastri@rice.edu ).

\subsection*{Materials availability}

This study did not generate new materials. The custom-fabricated dual-channel heat oscillator described in this work can be reproduced following the methods and specifications detailed in this paper.

\subsection*{Data and code availability}
\begin{itemize}
    \item All data reported in this paper will be shared by the lead contact upon reasonable request.
    \item All original COMSOL simulation files used in this study will be shared by the lead contact upon reasonable request. This paper does not report original code.
    \item Any additional information required to reanalyze the data reported in this paper is available from the lead contact upon request.
\end{itemize}

\section*{ACKNOWLEDGMENTS}


A.M.-O.\ acknowledges financial support from SECIHTI (Mexico, scholarship number 2021-000014-01EXTF-00140). N.\,H.\ acknowledges support from the Robert A.\ Welch Foundation under grant C-1220. A.\,A.\ acknowledges support from the Robert A.\ Welch Foundation under grant C-2224. N.J.H.\ and A.A.\ acknowledge funding support from the Department of Energy's Solar Desalination Prize. This material is based upon work supported by the National Science Foundation under Grant NSF 2346014.

\section*{AUTHOR CONTRIBUTIONS}


Q. Y. performed the numerical simulations, theoretical analysis and wrote the first draft of
the manuscript. A. M.-O. fabricated the experimental setup and performed the experiments.
W. S. contributed to the interpretation of the theoretical findings. G. W. contributed to the
thermal analysis of the system, the interpretation of the results and the contextualization of
the study into the broader energy storage landscape. N. H. supervised the execution of the
experiments and participated in coordinating the research. A. A. conceived the study,
coordinated the research and supervised the execution of the numerical calculations. All
authors discussed the research and participated in the writing of the manuscript.

\section*{DECLARATION OF INTERESTS}

A. A. and N. H. are listed as co-inventors in a patent, 'Resonant thermal oscillator to improve output of a thermo-fluidic system'.
A. A. owns a small share ($<$5 \%) in Localized Water Solutions, a water treatment technology company. He has no active role in it.

\section*{DECLARATION OF GENERATIVE AI AND AI-ASSISTED TECHNOLOGIES}

During the preparation of this work, the authors used generative AI tools (Anthropic Claude and OpenAI ChatGPT) to assist with language editing, phrasing, and improving the clarity of the manuscript text. These tools were not used to generate, analyze, or interpret any scientific data, simulations, or results. After using these tools, the authors reviewed and edited the content as needed and take full responsibility for the content of the publication.

\section*{SUPPLEMENTAL INFORMATION INDEX}

\begin{description}
  \item Document S1. Figures S1--S20 and Table S1, related to Figures 1--7
\end{description}




\bibliography{references}

@techreport{Avghad2016,
  author    = {Avghad, S. and Keche, A. and Kousal, A.},
  title     = {Thermal Energy Storage: A Review},
  year      = {2016},
}

@article{Sharma2025,
  author    = {Sharma, P. and Debnath, B. and Shukla, A. K. and Pawar, J. and Singh, G.},
  title     = {A Comprehensive Review of Sensible Heat Thermal Energy Storage for High Temperature Applications},
  journal   = {Energy Storage},
  year      = {2025},
  volume    = {7},
  number    = {5},
  pages     = {e70190},
  doi       = {10.1002/est2.70190},
}

@incollection{Raghav2020,
  author    = {Raghav, G. and Nagpal, M. and Kumar, S.},
  title     = {Performance Analysis of High Temperature Sensible Heat Solar Energy Storage System},
  editor    = {Singh, S. and Ramadesigan, V.},
  publisher = {Springer Singapore},
  year      = {2020},
  pages     = {387--396},
}

@article{Saleem2024,
  author    = {Saleem, A. and Ambreen, T. and Ugalde-Loo, C. E.},
  title     = {Energy Storage-Integrated Ground-Source Heat Pumps for Heating and Cooling Applications: A Systematic Review},
  journal   = {J. Energy Storage},
  year      = {2024},
  volume    = {102},
  pages     = {114097},
  doi       = {10.1016/j.est.2024.114097},
}

@book{Subekti2011,
  author    = {Subekti, N.},
  title     = {{IFIC} Book},
  year      = {2011},
}

@inproceedings{Reepmeyer2004,
  author    = {Reepmeyer, F. and Repke, J. and Forner, F. and Wozny, G.},
  title     = {How to Start Up Reactive Distillation Towers},
  year      = {2004},
}

@article{GomezRueda2022,
  author    = {Gomez-Rueda, Y. and Verougstraete, B. and Ranga, C. and Perez-Botella, E. and Reniers, F. and Denayer, J. F. M.},
  title     = {Rapid Temperature Swing Adsorption Using Microwave Regeneration for Carbon Capture},
  journal   = {Chem. Eng. J.},
  year      = {2022},
  volume    = {446},
  pages     = {137345},
  doi       = {10.1016/j.cej.2022.137345},
}

@article{Miralles2013,
  author    = {Miralles, V. and Huerre, A. and Malloggi, F. and Jullien, M.-C.},
  title     = {A Review of Heating and Temperature Control in Microfluidic Systems: Techniques and Applications},
  journal   = {Diagnostics},
  year      = {2013},
  volume    = {3},
  number    = {1},
  pages     = {33--67},
  doi       = {10.3390/diagnostics3010033},
}

@article{Luyben2019,
  author    = {Luyben, W. L.},
  title     = {Temperature Setpoint-Ramp Control Structure for Batch Reactors},
  journal   = {Chem. Eng. Sci.},
  year      = {2019},
  volume    = {208},
  pages     = {115124},
  doi       = {10.1016/j.ces.2019.07.042},
}

@techreport{Hurst2024,
  author    = {Hurst, K. E. and Springer, M. and Wikoff, H. and Cory, K. and Garfield, D. and Ruth, M. and {Bench Reese}, S.},
  title     = {Industrial Energy Storage Review},
  institution = {National Renewable Energy Laboratory (NREL), Golden, CO (United States)},
  number    = {NREL/TP--6A20-85634},
  year      = {2024},
  doi       = {10.2172/2473658},
}

@article{Woods2021,
  author    = {Woods, J. and Mahvi, A. and Goyal, A. and Kozubal, E. and Odukomaiya, A. and Jackson, R.},
  title     = {Rate Capability and {Ragone} Plots for Phase Change Thermal Energy Storage},
  journal   = {Nat. Energy},
  year      = {2021},
  volume    = {6},
  number    = {3},
  pages     = {295--302},
  doi       = {10.1038/s41560-020-00767-5},
}

@article{Wang2026,
  author    = {Wang, D. and Cao, P.-C. and Wang, Y. and Qi, M. and Ju, R. and Chen, H. and Qiu, C.-W. and Li, Y.},
  title     = {Scattering Symmetry of Diffusive Systems},
  journal   = {Phys. Rev. B},
  year      = {2026},
  doi       = {10.1103/b1rb-nmb3},
}

@article{Cao2024,
  author    = {Cao, P.-C. and Ju, R. and Wang, D. and Qi, M. and Liu, Y.-K. and Peng, Y.-G. and Chen, H. and Zhu, X.-F. and Li, Y.},
  title     = {Observation of Parity-Time Symmetry in Diffusive Systems},
  journal   = {Sci. Adv.},
  year      = {2024},
  volume    = {10},
  number    = {16},
  pages     = {eadn1746},
  doi       = {10.1126/sciadv.adn1746},
}

@article{Liu2024,
  author    = {Liu, J. and Xu, L. and Huang, J.},
  title     = {Spatiotemporal Diffusion Metamaterials: Theories and Applications},
  journal   = {Appl. Phys. Lett.},
  year      = {2024},
  volume    = {124},
  number    = {21},
  pages     = {210502},
  doi       = {10.1063/5.0208656},
}

@article{Yang2024,
  author    = {Yang, F. and Zhang, Z. and Xu, L. and Liu, Z. and Jin, P. and Zhuang, P. and Lei, M. and Liu, J. and Jiang, J.-H. and Ouyang, X. and Marchesoni, F. and Huang, J.},
  title     = {Controlling Mass and Energy Diffusion with Metamaterials},
  journal   = {Rev. Mod. Phys.},
  year      = {2024},
  volume    = {96},
  number    = {1},
  pages     = {015002},
  doi       = {10.1103/RevModPhys.96.015002},
}

@article{Li2021,
  author    = {Li, Y. and Li, W. and Han, T. and Zheng, X. and Li, J. and Li, B. and Fan, S. and Qiu, C.-W.},
  title     = {Transforming Heat Transfer with Thermal Metamaterials and Devices},
  journal   = {Nat. Rev. Mater.},
  year      = {2021},
  volume    = {6},
  number    = {6},
  pages     = {488--507},
  doi       = {10.1038/s41578-021-00283-2},
}

@article{Zhang2023,
  author    = {Zhang, Z. and Xu, L. and Qu, T. and Lei, M. and Lin, Z.-K. and Ouyang, X. and Jiang, J.-H. and Huang, J.},
  title     = {Diffusion Metamaterials},
  journal   = {Nat. Rev. Phys.},
  year      = {2023},
  volume    = {5},
  number    = {4},
  pages     = {218--235},
  doi       = {10.1038/s42254-023-00565-4},
}

@article{Li2019,
  author    = {Li, Y. and Peng, Y.-G. and Han, L. and Miri, M.-A. and Li, W. and Xiao, M. and Zhu, X.-F. and Zhao, J. and Al\`{u}, A. and Fan, S. and Qiu, C.-W.},
  title     = {Anti--Parity-Time Symmetry in Diffusive Systems},
  journal   = {Science},
  year      = {2019},
  volume    = {364},
  number    = {6436},
  pages     = {170--173},
  doi       = {10.1126/science.aaw6259},
}

@article{Ju2023,
  author    = {Ju, R. and others},
  title     = {Convective Thermal Metamaterials: Exploring High-Efficiency, Directional, and Wave-Like Heat Transfer},
  journal   = {Adv. Mater.},
  year      = {2023},
  doi       = {10.1002/adma.202209123},
}

@article{Alabastri2020,
  author    = {Alabastri, A.},
  title     = {Flow-Driven Resonant Energy Systems},
  journal   = {Phys. Rev. Appl.},
  year      = {2020},
  volume    = {14},
  number    = {3},
  pages     = {034045},
  doi       = {10.1103/PhysRevApplied.14.034045},
}

@article{Ye2023,
  author    = {Ye, Q. and Sanders, S. and Alabastri, A.},
  title     = {Resonant Energy Transfer and Storage in Coupled Flow-Driven Heat Oscillators},
  journal   = {PRX Energy},
  year      = {2023},
  volume    = {2},
  number    = {2},
  pages     = {023007},
  doi       = {10.1103/PRXEnergy.2.023007},
}

@article{Alabastri2020b,
  author    = {Alabastri, A. and Dongare, P. D. and Neumann, O. and Metz, J. and Adebiyi, I. and Nordlander, P. and Halas, N. J.},
  title     = {Resonant Energy Transfer Enhances Solar Thermal Desalination},
  journal   = {Energy Environ. Sci.},
  year      = {2020},
  volume    = {13},
  number    = {3},
  pages     = {968--976},
  doi       = {10.1039/c9ee03256h},
}

@article{Schmid2025,
  author    = {Schmid, W. and Machorro-Ortiz, A. and Ye, Q. and Nordlander, P. and Dongare, P. D. and Halas, N. J. and Alabastri, A.},
  title     = {Resonant Energy Transfer for Membrane-Free, Off-Grid Solar Thermal Humidification--Dehumidification Desalination},
  journal   = {Nat. Water},
  year      = {2025},
  volume    = {3},
  number    = {5},
  pages     = {605--616},
  doi       = {10.1038/s44221-025-00438-3},
}

@book{Svelto2013,
  author    = {Svelto, O. and Hanna, D. C.},
  title     = {Principles of Lasers},
  publisher = {Springer},
  year      = {2013},
}

@article{Wehmeyer2017,
  author    = {Wehmeyer, G. and Yabuki, T. and Monachon, C. and Wu, J. and Dames, C.},
  title     = {Thermal Diodes, Regulators, and Switches: Physical Mechanisms and Potential Applications},
  journal   = {Appl. Phys. Rev.},
  year      = {2017},
  volume    = {4},
  number    = {4},
  pages     = {041304},
  doi       = {10.1063/1.5001072},
}

@article{Ram2023,
  author    = {Ram, B. R. and Malik, V. and Naik, B. K. and Patel, K. S.},
  title     = {A Critical Review on Mechanical Heat Switches for Engineering and Space Applications},
  journal   = {Heat Transfer Eng.},
  year      = {2023},
  volume    = {44},
  number    = {19},
  pages     = {1789--1802},
  doi       = {10.1080/01457632.2022.2148348},
}

@article{Kommandur2023,
  author    = {Kommandur, S. and Kishore, R. A.},
  title     = {Contact-Based Passive Thermal Switch with a High Rectification Ratio},
  journal   = {ACS Eng. Au},
  year      = {2023},
  volume    = {3},
  number    = {2},
  pages     = {76--83},
  doi       = {10.1021/acsengineeringau.2c00046},
}

@article{Qi2022,
author = {Qi, Minghong and Wang, Dong and Cao, Pei-Chao and Zhu, Xue-Feng and Qiu, Cheng-Wei and Chen, Hongsheng and Li, Ying},
title = {Geometric Phase and Localized Heat Diffusion},
journal = {Advanced Materials},
volume = {34},
number = {32},
pages = {2202241},
doi = {https://doi.org/10.1002/adma.202202241},
url = {https://advanced.onlinelibrary.wiley.com/doi/abs/10.1002/adma.202202241},
eprint = {https://advanced.onlinelibrary.wiley.com/doi/pdf/10.1002/adma.202202241},
year = {2022}
}

@article{
Huberman2019,
author = {S. Huberman  and R. A. Duncan  and K. Chen  and B. Song  and V. Chiloyan  and Z. Ding  and A. A. Maznev  and G. Chen  and K. A. Nelson },
title = {Observation of second sound in graphite at temperatures above 100 K},
journal = {Science},
volume = {364},
number = {6438},
pages = {375-379},
year = {2019},
doi = {10.1126/science.aav3548},
URL = {https://www.science.org/doi/abs/10.1126/science.aav3548},
eprint = {https://www.science.org/doi/pdf/10.1126/science.aav3548}}

@article{Beaupere2018,
  author  = {Beaupere, No\'{e} and Soupremanien, Ulrich and Zalewski, Laurent},
  title   = {Nucleation triggering methods in supercooled phase change materials ({PCM}), a review},
  journal = {Thermochimica Acta},
  volume  = {670},
  pages   = {184--201},
  year    = {2018},
  doi     = {10.1016/j.tca.2018.10.009}
}

@article{Chen2023,
  author  = {Chen, Weiye and Chen, Lei and Li, Liangyu and Dong, Chuanshuai and  Zhang, Lizhi},
  title   = {Electrically-triggered nucleation of supercooled sodium acetate trihydrate phase change composites},
  journal = {Chemical Engineering Journal},
  volume  = {456},
  pages   = {141131},
  year    = {2023},
  doi     = {10.1016/j.cej.2022.141131}
}

@article{Wang2025AMPD,
  author  = {Wang, Yifei and Zhang, Yifan and Liu, Hong and Chen, Hui and Hu, Yang and Liu, Zhanjun},
  title   = {Supercooling Behavior of 2-Amino-2-methyl-1,3-propanediol for Thermal Energy Storage},
  journal = {Molecules},
  volume  = {30},
  number  = {10},
  pages   = {2206},
  year    = {2025},
  doi     = {10.3390/molecules30102206}
}

@article{McKay2013ThermalPulse,
  author  = {McKay, Ian Salomon and Wang, Evelyn N.},
  title   = {Thermal pulse energy harvesting},
  journal = {Energy},
  volume  = {57},
  pages   = {632-640},
  year    = {2013},
  doi     = {10.1016/j.energy.2013.05.045}
}

@article{Chen2015TEG,
  author  = {Chen, Leisheng and Lee, Jaeyoung},
  title   = {Effect of pulsed heat power on the thermal and electrical performances of a thermoelectric generator},
  journal = {Applied Energy},
  volume  = {150},
  pages   = {138--149},
  year    = {2015},
  doi     = {10.1016/j.apenergy.2015.04.009}
}

@techreport{Forsberg2017LWR,
  author      = {Forsberg, Charles and Haratyk, J. and Jenking, J. and Wooten, J. and Gasper, J. and Brick, S. and Varrin, R. and Schneider, E. and Mann, N. and Doster, M. and Stansbury, C. and Ding, Y. and Bindra, H. and McLauchlan, N. and Buscheck, T. and Lester, R. and Curtis, D. and Krall, T. and Sowder, A. and Jurewicz, J.},
  title       = {Light Water Reactor Heat Storage for Peak Power and Increased Revenue: Focused Workshop on Near-Term Options},
  institution = {Massachusetts Institute of Technology, MIT-ANP-TR-170},
  year        = {2017},
  address     = {Cambridge, MA}
}

@article{Wei2026MoltenSalt,
  author  = {Wei, Le and Fan, Bingfen and Zhang, Yi and Fang, F.},
  title   = {Flexible Peak Shaving Control Strategy of Thermal Power Unit Coupled With Molten Salt Heat Storage System},
  journal = {Proceedings of the CSEE},
  volume  = {46},
  number  = {2},
  pages   = {691--702},
  year    = {2026},
  doi     = {10.13334/j.0258-8013.pcsee.241471}
}

@article{Krishnan2005PCM,
  author  = {Krishnan, Shankar and Garimella, Suresh V. and Kang, Sung S.},
  title   = {A novel hybrid heat sink using phase change materials for transient thermal management of electronics},
  journal = {IEEE Transactions on Components and Packaging Technologies},
  volume  = {28},
  number  = {2},
  pages   = {281--289},
  year    = {2005},
  doi     = {10.1109/TCAPT.2005.848534}
}

@phdthesis{Tianyu2021ThermalSwitch,
  author = {Yang, Tianyu},
  title  = {Controlling heat transfer in electronic packaging using thermal switches and high power thermal buffers},
  school = {University of Illinois at Urbana-Champaign},
  year   = {2021}
}

@article{Shamberger2016,
  author  = {Shamberger, Patrick J.},
  title   = {Cooling Capacity Figure of Merit for Phase Change Materials},
  journal = {Journal of Heat Transfer},
  volume  = {138},
  number  = {2},
  pages   = {024502},
  year    = {2016},
  month   = feb,
  doi     = {10.1115/1.4031252},
  publisher = {ASME},
  url     = {https://asmedigitalcollection.asme.org/heattransfer/article-abstract/138/2/024502/384475}
}

@article{Zhu2022OscillatingGadolinium,
  author    = {Qing Zhu and Kaitlyn Zdrojewski and Lorenzo Castelli and Geoff Wehmeyer},
  title     = {Oscillating Gadolinium Thermal Diode Using Temperature-Dependent Magnetic Forces},
  journal   = {Advanced Functional Materials},
  year      = {2022},
  volume    = {32},
  number    = {43},
  pages     = {2206733},
  doi       = {10.1002/adfm.202206733},
  url       = {https://doi.org/10.1002/adfm.202206733}
}

@article{Miner2004,
  author  = {Miner, A. and Ghoshal, U.},
  title   = {Cooling of high-power-density microdevices using liquid metal coolants},
  journal = {Applied Physics Letters},
  volume  = {85},
  number  = {3},
  pages   = {506--508},
  year    = {2004},
  doi     = {10.1063/1.1772862}
}

@article{Deng2009,
  author  = {Deng, Yueguang and Liu, Jing},
  title   = {Corrosion development between liquid gallium and four typical metal substrates used in chip cooling device},
  journal = {Applied Physics A},
  volume  = {95},
  pages   = {907--915},
  year    = {2009},
  doi     = {10.1007/s00339-009-5098-1}
}

\newpage

\section*{MAIN FIGURE TITLES AND LEGENDS}




\noindent\includegraphics[width=0.85\linewidth]{Figure_1.pdf}
\subsection*{Figure 1. Comparison of approaches to transient heat generation from static thermal input}
Representative approaches mapped by estimated peak release-rate enhancement, $P_{\mathrm{out}}/P_{\mathrm{in}}$, and operating-point flexibility.
\newline
(A) Marker color indicates whether the input can remain on during release (yes, partial, or no/typically no).
\newline
(B) Marker shape indicates release predictability (deterministic on-demand, deterministic periodic, stochastic, or passive deterministic).
\newline
Active thermal Q-switching (this work) occupies a distinct region by combining deterministic on-demand release, sustained-input compatibility, and tunable power--energy trajectories within the same hardware. Values for prior methods (PCM thermal buffers, oscillating thermal switches, steam accumulators/molten-salt peak shaving, and triggered-release supercooled PCMs) are representative estimates compiled from the cited literature.
\newpage

\noindent\includegraphics[width=0.85\linewidth]{Figure_2_Update.pdf}
\subsection*{Figure 2. Q-switching for thermal power amplification: concept and simulations}
Schematic and simulated dynamics of a counter-flow heat oscillator under active Q-switching.
\newline
(A) Resonant configuration. Two countercurrent flow rates, $\dot{V}_1$ and $\dot{V}_2$, with density $\rho$, specific heat $c$, and thermal conductivity $k$, exchange heat through a thermal conductor at the interface (effective heat transfer coefficient $h_{\mathrm{int}}$) of channels with length $L$ and thickness $w$. A steady input power $P_{\mathrm{in}}$ enters the top surface of fluid 1. The two-channel system loses power partially to the environment, $\Gamma_{\mathrm{env}}$, and through the outlets, $P_{\mathrm{out}}$. When $\dot{V}_1/\dot{V}_2 = 1$, the system is at resonance, sustains large heat oscillations, accumulates thermal energy $E_{\mathrm{th}}$, and exhibits a large Q-factor. Circling red arrows represent heat flow in space; the harmonic red line shows the dynamics of heat flow between channels over time.
\newline
(B) Non-resonant configuration shortly after Q-switching. \textit{Left}: same as in (A), but with $\dot{V}_1/\dot{V}_2 \neq 1$; the system is off-resonance and previously accumulated thermal energy rapidly leaves the system. \textit{Right}: same as left, but the system releases energy by damping the inter-channel interaction (suppressed $h_{\mathrm{int}}$).
\newline
(C) Simulated accumulated thermal energy density $E_{\mathrm{th}}$ (black, left axis) and normalized outlet power $P_{\mathrm{out}}/P_{\mathrm{in}}$ (red, right axis) versus time for a system initially at resonance (accumulation phase, 10{,}000 s) and then switched to off-resonance (release phase, 100 s).
\newline
Simulation parameters: Galinstan-filled channels, $w = 1$ mm, $L = 1$ m, $u = 0.04$ m s$^{-1}$, $I_{\mathrm{in}} = 0.5$ kW m$^{-2}$, $h_{\mathrm{env}} = 1$ W m$^{-2}$ K$^{-1}$.
\newpage

\noindent\includegraphics[width=0.85\linewidth]{Figure_3.pdf}
\subsection*{Figure 3. Experimental configuration and illumination profile of the dual-channel microfluidic heat oscillator}
Fabricated water-based heat oscillator and characterization of the incident illumination.
\newline
(A) Three-dimensional schematic of the fabricated device, consisting of a 316L stainless steel midplane sandwiched between two polycarbonate slabs, forming two counter-flow water channels (20 cm wide $\times$ 50 cm long, 2 mm channel thickness). One side of the stainless-steel plate is coated with high-temperature black paint to enable efficient photothermal absorption. The system is illuminated from above by a high-power LED source positioned 4 cm above the device.
\newline
(B) Measured spatial irradiance distribution along the 50 cm channel length produced by the LED lamp, showing a stepped intensity profile arising from the lamp geometry. The irradiance varies from approximately 120 to 1570 W m$^{-2}$ across five distinct zones. This non-uniform illumination profile is incorporated as a spatially varying boundary condition into the comparison numerical simulation.
\newpage

\noindent\includegraphics[width=0.85\linewidth]{Figure_4.pdf}
\subsection*{Figure 4. Heat oscillator response to Q-switching by flow detuning: experiments and simulations}
Demonstration of active Q-switching via flow-rate detuning under continuous illumination.
\newline
(A) \textit{Top}: schematic of the flow-rate configurations for the experimental system, showing matched flows in the high-Q (resonant) state (left) and a faster Channel 1 in the low-Q (off-resonant) state (right).
\newline
(B) \textit{Middle}: imposed flow rates versus time for Channel 1 (black) and Channel 2 (red). Channel 1 is abruptly increased from 9.4 to 35.7 cm$^3$ min$^{-1}$ at $t = 5$ h, while Channel 2 is held at 8.9 cm$^3$ min$^{-1}$.
\newline
(C) \textit{Bottom}: experimental (dots) and simulated (solid lines) outlet temperature increase relative to ambient for Channel 1 (green) and Channel 2 (purple), left axis. Calculated outlet power for Channel 1, normalized to the pre-switch steady-state power $P_{\mathrm{steady}}$ (red, right axis). The shaded region marks the resonant accumulation phase; the unshaded region marks the off-resonant release phase. A $\sim$5-fold transient power amplification from Channel 1 is observed immediately after detuning.
\newline
Experimental uncertainties (standard deviation, $n = 2$ repetitions) are
typically smaller than the plotted symbols and are omitted for clarity;
the same flow-detuning protocol with error bars is shown in Fig.~S8
(Supplemental Information, Section~S5.3).
\newpage

\noindent\includegraphics[width=0.85\linewidth]{Figure_5.pdf}
\subsection*{Figure 5. Resonant system response to interchannel thermal interaction modulation: simulations}
Active Q-switching by suppression of interfacial thermal coupling in a water-based system; $L = 0.5$ m, $\dot{V}_1 = 32$ cm$^3$ min$^{-1}$,
illumination 0.1 kW m$^{-2}$, $h_{\mathrm{env}} = 1$ W m$^{-2}$ K$^{-1}$.
\newline
(A) High-Q accumulation: resonant coupling sustains heat recirculation
(green loop).
\newline
(B) Low-Q release: an insulating interfacial layer (red ``X''; $t = 0.1$
mm, $k = 10^{-5}$ W m$^{-1}$ K$^{-1}$) suppresses the coupling.
\newline
(C) Normalized powers $P_{\mathrm{out}}/P_{\mathrm{in}}$,
$\Gamma_{\mathrm{env}}/P_{\mathrm{in}}$, and
$(P_{\mathrm{out}}+\Gamma_{\mathrm{env}})/P_{\mathrm{in}}$, and average
energy density $\varepsilon_{\mathrm{avg}}$, versus time; switching at
$t = 6000$ s.
\newline
(D) Temperature maps and heat-flux vectors at 3800, 6208, and 7800 s.
Flow rates are constant throughout; only the interfacial coupling is
switched.

\newpage

\noindent\includegraphics[width=0.85\linewidth]{Figure_6.pdf}
\subsection*{Figure 6. Comparison between flow-detuning and interaction-modulation Q-switching approaches: simulations}
Equivalence of two independent Q-switching actuation modes in a Galinstan-filled system ($L = 1$ m, $w = 1$ mm, illumination 1 kW m$^{-2}$).
\newline
(A) Normalized outlet power $P_{\mathrm{out}}/P_{\mathrm{in}}$ during accumulation and subsequent release triggered by two control protocols: modulation of the interfacial conductivity $k(t)/k_0$ (green) and modulation of the flow velocity $u(t)/u_0$ (blue). \textit{Left inset}: imposed control signals for $k(t)/k_0$ and $u(t)/u_0$. \textit{Right inset}: zoom of the first switching event, showing that the imposed velocity program (orange, right axis) produces the same extraction spike and relaxation behavior as conductivity gating.
\newline
(B) Outlet temperature $T_{\mathrm{out}}$ (solid lines) and system-averaged temperature $T_{\mathrm{avg}}$ (dashed lines), where $T_{\mathrm{avg}}$ is obtained by averaging over the full channel cross-section and along the channel length. \textit{Left inset}: zoom highlighting the narrow $T_{\mathrm{out}}$ burst and subsequent monotonic relaxation of $T_{\mathrm{avg}}$. \textit{Right inset}: schematic illustrating the two independent control parameters, $u(t)$ and $k(t)$. The black curve depicts the spiral-like trajectory of a representative state point in phase space, evolving under the heat-flux vector field.
\newline
A second actuation at $t \approx 10^5$ s demonstrates repeatability. Throughout the accumulation phase, the system is held at the resonant flow condition shown in the subpanel; switching is triggered by modulating either $k(t)$ or $u(t)$ at the prescribed switching event.
\newpage

\noindent\includegraphics[width=0.85\linewidth]{Figure_7_Updated.pdf}
\subsection*{Figure 7. Ragone-type operating maps: dynamic trajectories of Q-switched heat oscillators versus static operating points of conventional TES (simulations and literature data)}
Ragone-type representation of release performance, comparing actively Q-switched heat oscillators to conventional thermal energy storage (TES) media.
\newline
(A) Mass-normalized power density 
\newline (B) Normalized discharge power (average extracted power during release, normalized by the incident input power) plotted as a function of area-normalized energy density, highlighting performance scaling with device footprint. 
\newline Colored curves: simulated trajectories for water- and Galinstan-based single (2-CH) and vertically stacked (8-CH) oscillators; each point corresponds to a full accumulation--release cycle, with stored energy and released power defined in Eqs.~\ref{eq:deltaE}--\ref{eq:E90}. Marker color and the adjacent numeric label indicate the average outlet temperature during discharge (in $^\circ$C; separate color scales for water and Galinstan). Solid (dashed) curve segments correspond to system temperatures below (exceeding) the boiling point of the working fluid. Dotted curves: rate-capability references for representative PCM storage media (water/ice, tetradecane/graphite composite, $n$-tetradecane), compiled from the cited literature.\cite{Woods2021} Vertical dashed lines (stars) denote the sensible energy density required to heat water and Galinstan to their boiling points, excluding latent heat. 
\newline Illumination intensities: 0.28 kW m$^{-2}$ for water-based systems; 32 kW m$^{-2}$ for Galinstan-based systems.
\newpage

\end{document}